# Search versus Decision for Election Manipulation Problems*

Edith Hemaspaandra†
Department of Computer Science
Rochester Institute of Technology
Rochester, NY 14623, USA

Lane A. Hemaspaandra‡
Department of Computer Science
University of Rochester
Rochester, NY 14627, USA

Curtis Menton§
Department of Computer Science
University of Rochester
Rochester, NY 14627, USA

February 29, 2012, revised March 6, 2012

**Abstract**

Most theoretical definitions about the complexity of manipulating elections focus on the decision problem of recognizing which instances can be successfully manipulated, rather than the search problem of finding the successful manipulative actions. Since the latter is a far more natural goal for manipulators, that definitional focus may be misguided if these two complexities can differ. Our main result is that they probably do differ: If integer factoring is hard, then for election manipulation, election bribery, and some types of election control, there are election systems for which recognizing which instances can be successfully manipulated is in polynomial time but producing the successful manipulations cannot be done in polynomial time.

## 1 Introduction

Elections are such a ubiquitous model for human and electronic collective decision-making—and during the past few decades, with the rise of computers, multiagent systems, and the internet, elections have become important even in many "modern" challenges such collaborative filtering/recommender systems [GMHS99], planning [ER97], and reducing web spam [DKNS01]—that much work has been devoted to studying how to manipulate elections. However, the broad stream of theoretical work on the computational complexity of manipulative attacks on elections (see the surveys [FHHR09b, FHH10]) is largely centered on the complexity of the decision versions: Given an instance, determining whether there exists a successful manipulation (typically, ensuring that a given candidate wins, or ensuring that a given candidate does not win) of the given sort.

As a running example that we will use in this introduction, consider unweighted noncoalition manipulation, which was central in one of the seminal papers on manipulation ([BO91], see

*Also appears as URCS-TR-2012-971.

†Supported in part by grant NSF-CCF-1101452 and a Friedrich Wilhelm Bessel Research Award. Work done in part while visiting Heinrich-Heine-Universität Düsseldorf.

‡Supported in part by grants NSF-CCF-{0426761,0915792,1101479} and arc-dp110101792, and a Friedrich Wilhelm Bessel Research Award. Work done in part while visiting Heinrich-Heine-Universität Düsseldorf.

§Supported in part by grants NSF-CCF-{0915792,1101479}. Work done in part while visiting Heinrich-Heine-Universität Düsseldorf.

also [BTT89a]). For this problem, relative to some fixed election system, the inputs are the candidate set, the voter set consisting of a collection of nonmanipulative voters (whose preferences are each typically expressed by each voter as a preference ballot, e.g., Gore > Nader > Bush), and a single manipulative voter who has not yet set her vote but who has a "preferred" candidate $p$. And the question is: Does there exist a preference (vote) the manipulative voter can cast that will make $p$ win the election? This is typically viewed as a decision (language) problem, namely, as the set of all instances for which the answer to that question is "Yes."

Of course, what a manipulator might most want is not to know a successful manipulation *exists* (a decision problem), but rather to know what *specific action* (what vote, bribe, etc.) to take to achieve success (a search problem). For the case of our unweighted noncoalition manipulation example, the search version would be a function that takes the same input as the decision version but then either outputs that no successful strategic vote for the manipulative voter exists or, if a successful vote does exist, *outputs a successful vote—one that makes $p$ win.*

This paper studies whether these two goals' achievability can differ: whether decision versions of election problems can be easy yet their search versions intractable.

Virtually all papers in this area, to prove polynomial-time results for deciding when manipulative actions can succeed, actually give polynomial-time algorithms to produce the successful action. So one might suspect that perhaps that is always the case. For manipulation, bribery, and some types of control, we prove otherwise, under a complexity-theoretic hypothesis that is widely believed true. Our main contributions are:

- If $\mathrm{P} \neq \mathrm{NP} \cap \mathrm{coNP}$, then for each of manipulation (including in particular the case of our running example, unweighted noncoalition manipulation), bribery, and certain types of partition-control, there exist election systems for which there are polynomial-time algorithms to determine whether each given instance has a successful manipulative action, but no polynomial-time algorithm can exist that given an instance that is manipulable provides the successful manipulation. (It is widely believed in cryptography that integer factoring is hard. It is well known that if integer factoring is hard then $\mathrm{P} \neq \mathrm{NP} \cap \mathrm{coNP}$.) Informally put, the situation is that the frustrated world of polynomial-time computation will have to say things such as, "I can totally guarantee you that there are strategic votes you can cast to make Barack Obama win in the given electoral setting, but I have no idea what those votes are." We show that this bizarre setting can even occur in extremely simple cases, such as unweighted noncoalition (i.e., where we have just a single manipulative voter) manipulation. It follows immediately from our results that if $\mathrm{P} \neq \mathrm{NP} \cap \mathrm{coNP}$ then for each of the above-mentioned manipulative actions there exists an election system in which the search problem does not polynomial-time Turing reduce to the decision problem.

- In contrast, we show that for all the standard types of election-control actions based on adding or deleting voters or candidates, and for some of the standard election-control actions based on partitioning, the search problem (finding how to succeed) polynomial-time Turing reduces to the decision problem (knowing when one can succeed). It follows that, for these manipulative actions and for every election system, the bizarre type of behavior mentioned earlier cannot occur: Easy recognition of instances where success is possible implies polynomial-time algorithms for how to achieve success. While proving this, we notice that two pairs of control attacks assumed to differ in fact are identical problems, namely, for every election system, destructive control by partition of candidates and destructive control by run-off partition of candidates are the same set in both the standard tie-breaking cases (ties-eliminate and ties-promote); this reduces by two the number of distinct, standard control types.

- Regarding the results of the first bullet point above, which, when $\mathrm{P} \neq \mathrm{NP} \cap \mathrm{coNP}$ make decision easy but search hard, one might worry that search may be only infrequently hard. We address this by, as Theorems 5.1 and 5.3 and Corollaries 5.2 and 5.4, constructing manipulative-action problems whose search versions are *just as often hard as are those problems in* $\mathrm{NP} \cap \mathrm{coNP}$ *that have the highest density of hardness*, give or take a slight degree of flexibility. These are a novel results proving a transference of density-of-hardness from a class to a particular type of concrete problem.

Given that $\mathrm{P} \neq \mathrm{NP} \cap \mathrm{coNP}$ suffices to for some systems make the natural problem to care about (the search version) hard even as the problem that has been the theoretical literature's central definition (the decision version) declares the problem easy, we suggest that in definition and problem framing it may now be good to more energetically stress the importance of the search versions of election manipulation problems.

## 2 Preliminaries

### Elections

An election will consist of a set $C$ of candidates and a set $V$ of voter names-preferences. Each candidate is identified by her name, and since it is a set, no two candidates may have the same name. Each voter has a name, though in our constructions we will never use those names. Each voter has a preference over all the candidates (it is legal for different voters to have the same preferences). For all the cases discussed in this paper, the preference will be a tie-free linear ordering of the candidates.[1] We assume each vote is input distinctly (i.e., the voters' preferences come in as separate ballots; but it would be cheating for us to use the order of that input list within our proofs). So a typical election might be $C = \{$"Alice", "Bob", "Carol"$\}$ and $V = \{($"Joe", "Carol" $>$ "Bob" $>$ "Alice"$), ($"Bob", "Bob" $>$ "Carol" $>$ "Alice"$), ($"Frank", "Bob" $>$ "Carol" $>$ "Alice"$)\}$. In that example we used alphabetic strings as names, but actually we view the universe of legal names as being all binary strings (which one might or might not choose to relate to alphabetic strings).

Although each voter has a name, particular election systems are free to ignore the names in their actions, e.g., most familiar election systems, such as plurality-rule elections, ignore the names. For this reason, many earlier papers treat the vote set as if the preferences come in without any voter names. But that is jumping a step. The natural social-choice model is that voters have names, and the election system may access and use those, although it may choose not to, and indeed notions exist—e.g., anonymity (see [AB00] for a formal definition)—that formalize various notions of detachment from naming details. Our constructed systems are all anonymous.

Election systems, or voting systems, map from an election instance $(C, V)$ to a set of winners (i.e., to a set $W$, $\emptyset \subseteq W \subseteq C$[2]).

[1] Other approaches to preference expression, such as approval vectors, exist. But our "search reduces to decision" results will hold for any reasonable approach. Our "search is harder than decision when $\mathrm{P} \neq \mathrm{NP} \cap \mathrm{coNP}$" results will use the tie-free linear ordering model for concreteness, but in most or all cases, one could vary the proofs and their codings to address, for example, the approval-vector-as-vote-format case.

[2] Pure social choice papers often definitionally exclude the case $W = \emptyset$, but like most papers on computational social choice we allow it. It is a natural case, is symmetric with $W = C$, and in some elections it is an appropriate outcome. Let us give a real-world example of when having no winners can be an appropriate election outcome. To get into the Baseball Hall of Fame one needs to be approved by at least 75 percent of those members of the Baseball Writers' Association of America who vote, each voting for up to ten of the few-dozen nominated candidates (this example happens to be for a case in which votes are cast in an approval, rather than a preference-order, model). If

### Election Winner, Manipulation, Bribery, and Control Problems

For each fixed election system $\mathcal{E}$, one can define the election winner problem as follows (see [BTT89b]).

**Name:** $\mathcal{E}$-winner, or the winner problem for $\mathcal{E}$.

**Given:** Election $(C, V)$ and candidate $p \in C$.

**Question:** Is $p$ a winner of the election $(C, V)$ under election system $\mathcal{E}$?

This is actually, in the way universally accepted in computer science (see [GJ79]), describing a set, i.e., a language. That set is the set of all triples $\langle C, V, p \rangle$ such that the answer to the question is "Yes."[3]

We now briefly present the key definitions for the three most commonly studied types of manipulative actions: manipulation, bribery, and control. These three types were first studied, respectively, by Bartholdi, Orlin, Tovey, and Trick [BTT89a,BO91], Faliszewski, Hemaspaandra, and Hemaspaandra [FHH09], and Bartholdi, Tovey, and Trick [BTT92], for the "constructive" cases, i.e., where the goal is to make a particular candidate be a winner.[4] The "destructive" cases, where the goal is to ensure that a particular candidate is not a winner, were introduced by Conitzer, Sandholm, and Lang [CSL07] for manipulation, by Faliszewski, Hemaspaandra, and Hemaspaandra for bribery [FHH09], and by Hemaspaandra, Hemaspaandra, and Rothe for control [HHR07].

The manipulation problem is defined as follows, and models whether a coalition of strategic voters can make a certain candidate win.

**Name:** $\mathcal{E}$-unweighted coalition manipulation, or the unweighted coalition manipulation problem for $\mathcal{E}$; for short, the manipulation problem for $\mathcal{E}$.

**Given:** Candidate set $C$, nonmanipulative voter set $V_1$ (as a set of (name, preferences) ballots with preferences over $C$), manipulative voter set $V_2$ (as a set of names of the manipulative voters, none of whose names may belong to $V_1$), and a candidate $p \in C$.

**Question:** Is there some choice of preferences for the manipulative voters such that $p$ is a winner in the election in system $\mathcal{E}$ with candidates $C$ and voters $V_1 \cup V_2$?

This again is a decision problem consisting of the set of all inputs yielding the answer "Yes." However, there is a very natural search problem associated with this, which we will call manipulation

---

no one is so strong as to receive the support of 75 percent of the voters, then no one is inducted into the Baseball Hall of Fame that year. Indeed, in three years—1945, 1950, and 1960—there was an election in which no one made it to the 75 percent threshold, and so no one was inducted in those years.

[3]Implicit here is that some appropriate coding scheme and pairing function are used to encode the inputs and to combine them into a single string, where "appropriate" carries a variety of restrictions on the complexity of coding and decoding. However, in this paper we need nothing nonroutine regarding pairing and so do not discuss that. And when we use names in particularly interesting ways, we always make clear what those ways are.

[4]We say "be a winner" as this entire paper will focus on that notion, known as the nonunique-winner model. That model has broadly been the one previous papers favored. The exception is that the seminal paper on control [BTT92] used the unique-winner model, i.e., it asked whether one could make a given candidate become the one and only winner. However, perhaps because the unique-winner model in some sense is forcing a tie-breaking issue into the winner model of control, and also because manipulation is almost always studied in the nonunique-winner model, recent papers on control generally focus on the nonunique-winner model, though they sometimes add lengthy extra text to also cover the unique-winner model in all their results. We feel the nonunique-winner model is clearly the more natural model, and so we take the plunge and hold to it throughout this paper—give or take a few asides. That is, all our results and discussion are about the nonunique-winner model unless they explicitly say otherwise.

search, i.e., finding the successful action. In particular, a function $f$ solves the manipulation search problem (for a given election system) if on all inputs where the question's answer is "No" (i.e., all inputs not in the set that is the decision version) the function indicates in some clear way (e.g., by outputting -1) that manipulation is not possible, and on each input that belongs to the decision version, $f$ specifies settings to the preferences of the manipulative voters in such a way that those result in $p$ being a winner in the election $(C, V_1 \cup V_2)$. If some solution for the manipulation search problem is a polynomial-time computable function, we will say that the manipulation search problem is polynomial-time computable (or, in a slight abuse of notation, we will say that the manipulation search problem is in FP, the class of polynomial-time computable functions—the slight abuse is that what we really mean is that some solution to the problem is in FP).

One can also define "weighted" coalition manipulation, where each manipulative and nonmanipulative voter has a weight (how many times her vote counts). Our results on manipulation all will hold for that case too (as will be clear from the proofs). But it is more interesting that the results hold even in the unweighted case—and indeed, we will show they hold even when the number of manipulative voters is limited to being at most one.

Unlike manipulation, in bribery all voters have initial preferences. In the simplest model of bribery, voters are unweighted and each has unit cost to bribe. (By varying these parameters, the original paper [FHH09] obtained three other models: unweighted, priced; weighted, unpriced; and weighted, priced. Our results on bribery hold in all four models, essentially by the same proofs we give, but are cleanest and most striking in the simplest model, and so we present them in that model, and define just that model.) This problem models whether having the ability to reshape (bribe) the preferences of a number of voters allows one to make a given candidate win.

**Name:** $\mathcal{E}$-bribery, or the bribery problem for $\mathcal{E}$.

**Given:** Election $(C, V)$, candidate $p \in C$, and integer $b \geq 0$.

**Question:** Does there exist some collection of at most $b$ voters, and a way of setting their votes, so that in the election under $\mathcal{E}$ in which those votes are thus set and the other voters cast the votes the input specified for them, $p$ is a winner?

Again, this is and should be viewed as a decision problem—as a set. It has a natural search version, which we will call bribery search. That version (or if one wants to be very formal, a solution for that version) of course will (correctly) say either that no successful bribe is possible or will give a successful bribe (i.e, it will specify the new preferences for a set of at most $b$ voters so as to make $p$ be a winner). We will speak of this being in polynomial time and/or FP in the analogous way as was described regarding manipulation.

Finally, we come to election control, the most varied, the most difficult to describe, but in our opinion the most interesting of the three most studied types of manipulative attacks on elections. Control asks whether by various adjustments to the participation and structure of an election, a given candidate can be made a winner. A natural set of control actions was specified in the seminal paper of Bartholdi, Tovey, and Trick on control [BTT92], and we adopt that set, very slightly modified—as is now done in most papers—to treat adding of candidates symmetrically with the other add/delete types (as suggested by [FHHR09a]) and to be clear in the "partition" cases about how first-round ties are handled (following [HHR07]). Those control types are adding candidates, deleting candidates, adding voters, deleting voters, partition of voters, run-off partition of candidates, and partition of candidates. These loosely model many real-life settings, ranging from get-out-the-vote drives to voter suppression to having a culling "primary" round to encouraging (or discouraging) "spoiler" candidates (see for example [FHHR09a] for extended discussions of how these model various real-life scenarios). Each of the three partition control types is actually two control types—one (denoted by a TP—"ties promote"—modifier) for the model in which if a first-round election has multiple winners they all move forward to the second round, and one (denoted by a TE—"ties eliminate"—modifier) for the model in which one moves forward from a first-round election only if one is the unique winner of that contest. We mention is passing that if in one's definition of the control question one is asking (as is the case in the present paper) "can $p$ be made a winner?", i.e., the "nonunique-winner model," then we feel the only reasonably partition types are the three TP ones, and if one's control question is "can $p$ be made a unique winner?", i.e., the "unique-winner model" (which is not the model used in the present paper), then we feel the only reasonably partition types are the three TE ones. That is, tie-handling in the first round and in the control question (which in partition cases is in effect controlling how second-round ties affect things) should be harmonized. However, that is merely our recommendation, and in this paper we follow the field's current norm of trying to explore and understand both the TP and TE cases, even though we are ourselves firmly in the nonunique-winner model.

We will now specify the standard control types, but rather succinctly as there are so many, rather than using Name/Given/Question for each. For consistency, we follow in many of these cases word-for-word the definitions given in earlier papers, especially [FHHR09a,FHH11a].

**Definition 2.1** *Let $\mathcal{E}$ be an election system.*

*(a) In the control by adding candidates problem for $\mathcal{E}$ we are given two disjoint sets of candidates $C$ and $A$, $V$ a collection of voters and their preferences over $C \cup A$, a candidate $p \in C$, and a nonnegative integer $K$. We ask if there is a set $A' \subseteq A$ such that (a) $\|A'\| \leq K$, and (b) $p$ is a winner of $\mathcal{E}$ election $(C \cup A', V)$.*[5]

*(b) In the control by deleting candidates problem for $\mathcal{E}$ we are given an election $(C, V)$, a candidate $p \in C$, and a nonnegative integer $K$. We ask if there is a set $C' \subseteq C$ such that (a) $\|C'\| \leq K$, (b) $p \notin C'$, and (c) $p$ is a winner of $\mathcal{E}$ election $(C - C', V)$.*

*(c) In the control by adding voters problem for $\mathcal{E}$ we are given a set of candidates $C$, two collections of voters (and their preferences), $V$ and $W$, with the preferences being over $C$, a candidate $p \in C$, and a nonnegative integer $K$. We ask if there is a subcollection $W' \subseteq W$ such that (a) $\|W'\| \leq K$, and (b) $p$ is a winner of $\mathcal{E}$ election $(C, V \cup W')$.*

*(d) In the control by deleting voters problem for $\mathcal{E}$ we are given an election $(C, V)$, a candidate $p \in C$, and a nonnegative integer $K$. We ask if there is a collection $V'$ of voters that can be obtained from $V$ be deleting at most $K$ voters such that $p$ is a winner of $\mathcal{E}$ election $(C, V')$.*

*(e) In the control by partition of voters problem for $\mathcal{E}$, in the TP or TE tie-handling rule model, we are given an election $(C, V)$ and a candidate $p \in C$. Is there a partition[6] of $V$ into $V_1$ and $V_2$ such that $p$ is a winner of the two-stage election where the winners of election*

[5]It is very important to note that throughout the definitions of all types of control, and throughout this paper, when we speak of an election, $(C', V')$, we always implicitly mean that each vote in $V'$ is passed to the election system only as the version of itself restricted to the candidates in $C'$. For example, in the present subpart of the definition, each vote in $V$, even though in the input being a name and a preference over all the candidates in $C \cup A$, is passed to the election system just as a name and a preference over the candidates in $C \cup A'$. This natural restriction of the votes to just their parts related to the candidates in the elections also applies in the other candidate control types, namely, deletion of candidates and all the candidate partition types. This is the normal approach in defining control types, but we stress it because if we did not follow this approach, we could cheat in some of our constructions and use parts of a vote regarding candidates not in the election to pass/control information.

[6]A partition of a set $A$ is a pair $(A_1, A_2)$ such that $A_1 \cup A_2 = A$ and $A_1 \cap A_2 = \emptyset$.

*$(C, V_1)$ that survive the tie-handling rule compete against the winners of $(C, V_2)$ that survive the tie-handling rule? Each subelection (in both stages) is conducted using election system $\mathcal{E}$.*

(f) *In the control by run-off partition of candidates problem for $\mathcal{E}$, in the TP or TE tie-handling rule model, we are given an election $(C, V)$ and a candidate $p \in C$. Is there a partition of $C$ into $C_1$ and $C_2$ such that $p$ is a winner of the two-stage election where the winners of subelection $(C_1, V)$ that survive the tie-handling rule compete against the winners of subelection $(C_2, V)$ that survive the tie-handling rule? Each subelection (in both stages) is conducted using election system $\mathcal{E}$.*

(g) *In the control by partition of candidates problem for $\mathcal{E}$, in the TP or TE tie-handling rule model, we are given an election $(C, V)$ and a candidate $p \in C$. Is there a partition of $C$ into $C_1$ and $C_2$ such that $p$ is a winner of the two-stage election where the winners of subelection $(C_1, V)$ that survive the tie-handling rule compete against all candidates in $C_2$? Each subelection (in both stages) is conducted using election system $\mathcal{E}$.*

(h) *In the control by unlimited adding candidates problem for $\mathcal{E}$ we are given two disjoint sets of candidates $C$ and $A$, $V$ a collection of voters and their preferences over $C \cup A$, and a candidate $p \in C$. We ask if there is a set $A' \subseteq A$ such that $p$ is a winner of $\mathcal{E}$ election $(C \cup A', V)$. (This probably should be viewed as a "deprecated" control type. It was the version of adding candidates that was included in the seminal control paper of Bartholdi, Tovey, and Trick [BTT92], but its lack of a bound on the number of additions was asymmetric with their three other addition/deletion control types, and so subsequent papers have focused instead on the more symmetric case that is listed above under the name "adding candidates." However, since this control type is part of the history of the problem, we also include in this paper a statement of what holds for this case.)*

Again, those are all decision problems, i.e., sets. And they have the obvious search versions, which we will refer to in ways analogous to those we mentioned earlier regarding manipulation and bribery.

All the manipulation, bribery, and control problems defined so far are about trying to make a certain candidate be a winner. We will henceforward when mentioning these problems always add the word "constructive," to indicate that the problem is about making the specified candidate be a winner. As alluded to earlier, for every problem we have defined there is a "destructive" version, where the question is whether one can ensure that the specified candidate is not a winner. Both the constructive and destructive problems have both decision and search versions, in the obvious way.

## Polynomial-Time Reductions

Polynomial-time many-one reductions ($\leq_m^p$) are defined by $A \leq_m^p B$ if there exists a polynomial-time function $\sigma$ such that $(\forall x)[x \in A \Leftrightarrow \sigma(x) \in B]$.

A (decision or search) problem $A$ is said to polynomial-time Turing reduce ($\leq_T^p$-reduce) to a decision problem $B$ if there is a machine $M$ such that (a) $M^B$ runs in polynomial time (relative to the length of its input), and (b) if $A$ is a decision problem then the language accepted by $M^B$ is $A$, and if $A$ is a search problem then $M^B$ computes a function that is a solution of the search problem. Here, $M^B$ means machine $M$ given a unit-cost subroutine testing membership in $B$ (i.e., $M$ has a "black-box" or "oracle" for $B$); see [HU79]—this is the standard definition of polynomial-time Turing reductions, which along with polynomial-time many-one reductions are the central ways theoretical computer science links and compares the complexity of problems. For example, if we say that $\mathcal{E}$-manipulation search polynomial-time Turing reduces to $\mathcal{E}$-manipulation, that means that given an instance of the $\mathcal{E}$-manipulation problem (but being interested in getting an action, i.e., we are doing the search version), we can in polynomial time, given access to an oracle for the set $\mathcal{E}$-manipulation, correctly either state that successful manipulation is impossible or output a successful manipulation. Since all Turing reductions in this paper will be polynomial-time Turing reductions, we will sometimes omit the words "polynomial-time" before "Turing reduces/reduction."

We move directly on to the presentation of our results, and then provide a discussion of related work.

## 3 Results

The tightly related goals of this paper are to determine for which manipulative actions

(a) for all election systems, search (polynomial-time Turing) reduces to decision,

and to determine for which manipulative actions

(b) there exists some election system $\mathcal{E}$, whose winner problem is in P, for which the decision version of the manipulative action is in P yet the search version of the decision problem is not polynomial-time computable (i.e., no FP function solves the search version—which we will, in a slight abuse of formalism, feel free to express by the phrase "the search version is not in FP").

These are related, as "(a)" clearly implies "NOT (b)."

For manipulation, bribery, and every standard type of control, we in effect strongly resolve this. That is, for some, we prove (a)—which of course implies NOT (b) (in fact, it implies even that "NOT (b′)," where (b′) is (b) with the "winner problem in P" requirement removed). And for all the others we prove, under the complexity-theoretic assumption $\mathrm{P} \neq \mathrm{NP} \cap \mathrm{coNP}$, that (b) holds—which of course implies NOT (a). Of course, the more striking group of cases is the latter collection—manipulative actions for which for some election system with an easy (i.e., polynomial-time) winner problem we can easily (i.e., in polynomial time) for a given setting determine whether a successful attack exists, and yet there can exist no polynomial-time algorithm to always tell us what the successful attack action (that we know exists!) is.

In the process of proving the latter group of cases—the cases where for the particular manipulative action we show $\mathrm{P} \neq \mathrm{NP} \cap \mathrm{coNP}$ implies (b)—we will do even more than promised above. We will not only show that $\mathrm{P} \neq \mathrm{NP} \cap \mathrm{coNP}$ implies (b), but we will characterize (b), for each of those manipulative actions, as being equivalent to the right-hand side condition of the so-called Borodin-Demers Theorem from computational complexity theory. It follows that not only does $\mathrm{P} \neq \mathrm{NP} \cap \mathrm{coNP}$ imply (b) for all of the cases we address regarding that, but also all those cases stand or fall as one: Although we need a rich variety of complex election schemes and tricky coding schemes to prove our results, from those results and that work we establish that twelve different instances of whether (b) holds are all, under their surface, the same issue.

In the process of proving the other group of cases—the cases where for the particular manipulative action it holds that, for all election systems, search polynomial-time Turing reduces to decision—we will note, essentially finding this implicit but unnoticed in four characterizations already in the literature, that two pairs of control types that have always been viewed as distinct in fact pairwise collapse: viewed as sets, they are the exact same set. So all previous papers that gave separate proofs for the two elements of a collapsing pair were, unbeknownst to the authors, proving the same result twice.[7] The collapsing pairs are the following: DC-RPC-TP = DC-PC-TP (i.e., viewed as decision problems, destructive control by run-off partition of candidates in the ties-promote model is *exactly* the same problem—the same set of strings—as is destructive control by partition of candidates in the ties-promote model) and DC-RPC-TE = DC-PC-TE.

| Manipulative Action | Constructive | Destructive |
| --- | --- | --- |
| bribery | S $\not\leq$ D | S $\not\leq$ D |
| control by adding voters | S $\leq$ D | S $\leq$ D |
| control by deleting voters | S $\leq$ D | S $\leq$ D |
| control by partition of voters, ties promote | S $\not\leq$ D | S $\not\leq$ D |
| control by partition of voters, ties eliminate | S $\not\leq$ D | S $\not\leq$ D |
| control by adding candidates | S $\leq$ D | S $\leq$ D |
| control by deleting candidates | S $\leq$ D | S $\leq$ D |
| control by partition of candidates, ties promote | S $\not\leq$ D | S $\leq$ D |
| control by partition of candidates, ties eliminate | S $\not\leq$ D | S $\leq$ D |
| control by run-off partition of candidates, ties promote | S $\not\leq$ D | S $\leq$ D |
| control by run-off partition of candidates, ties eliminate | S $\not\leq$ D | S $\leq$ D |
| control by unlimited adding of candidates (note: deprecated type) | S $\leq$ D | S $\leq$ D |
| manipulation | S $\not\leq$ D | S $\not\leq$ D |

Table 1: **Results summary.** Key: "S $\leq$ D" is shorthand for: For each election system $\mathcal{E}$, the named constructive or destructive manipulative action has the property that its search version polynomial-time Turing reduces to its decision version. (Note that this implies that it is impossible for its decision version to be polynomial-time computable but its search version not to be polynomial-time computable.) "S $\not\leq$ D" is shorthand for: If $\mathrm{P} \neq \mathrm{NP} \cap \mathrm{coNP}$, then there exists an election system $\mathcal{E}$, having a polynomial-time winner problem, such that the named constructive or destructive manipulative action's decision problem is in polynomial time but its search problem is not in polynomial time. (Note that this implies that if $\mathrm{P} \neq \mathrm{NP} \cap \mathrm{coNP}$, then there is an election system $\mathcal{E}$ such that for the named constructive or destructive manipulative action, search does not polynomial-time Turing reduce to decision.)

Although we view proving our results as very important, we realize that not all readers will want or need the proof details, but rather may want to in an accessible, clear way know what holds, and to at least be shown the general idea/flavor of the proof approaches. So we structure the rest of this section as follows. We now give, as Table 1, an overview of our results. Section 3.1 presents our results about $\mathrm{P} \neq \mathrm{NP} \cap \mathrm{coNP}$ implying that there are simple election systems for which the decision version of a certain manipulative action is in polynomial time but the search version is not. That section also gives an informal, high-level view of our general proof approach for those results, using a particular case as an example. Section 3.2 presents our results showing that for many manipulative actions, search Turing reduces to decision, and describes the proof approach of those results, using two concrete cases as examples. Finally, Section 4 provides proofs of our results; readers not interested in proofs may wish to skip this section. We do, however, even in that section try to for most proofs give a short, high-level description of what goes on in the proof, typically either just before the proof or at the start of the proof.

[7] Although the present paper is in the nonunique winner model, to be fair to earlier papers it is important to note here that of the two pairs that we prove collapse (in the nonunique winner model), only one of those pairs collapses in the unique winner model. That itself is also a new result, and we prove it in Footnote 11.

## 3.1 Cases When the Manipulative-Action Decision Problem Is Easy but Its Search Problem Is Hard

Our main result, showing that if $\mathrm{P} \neq \mathrm{NP} \cap \mathrm{coNP}$ then there are easy election systems (i.e., having a polynomial-time winner problem) whose manipulative-action decision problem is easy but whose manipulative-action search problem is hard, is the following.

**Theorem 3.1** *If* $\mathrm{P} \neq \mathrm{NP} \cap \mathrm{coNP}$*, then for each manipulative action* $\mathcal{A}$ *belonging to the following list:*

1. *constructive manipulation,*
2. *destructive manipulation,*
3. *constructive bribery,*
4. *destructive bribery,*
5. *constructive control by partition of voters, ties promote,*
6. *destructive control by partition of voters, ties promote,*
7. *constructive control by partition of voters, ties eliminate,*
8. *destructive control by partition of voters, ties eliminate,*
9. *constructive control by partition of candidates, ties promote,*
10. *constructive control by partition of candidates, ties eliminate,*
11. *constructive control by run-off partition of candidates, ties promote, and*
12. *constructive control by run-off partition of candidates, ties eliminate,*

*there exists an election system* $\mathcal{E}$ *(which may differ based on* $\mathcal{A}$*), whose winner problem is in polynomial time, such that the* $\mathcal{A}$*-decision problem for* $\mathcal{E}$ *is in* P *but the* $\mathcal{A}$*-search problem for* $\mathcal{E}$ *is not polynomial-time computable.*

**Corollary 3.2** *If* $\mathrm{P} \neq \mathrm{NP} \cap \mathrm{coNP}$*, then for each of the manipulative actions* $\mathcal{A}$ *listed in Theorem 3.1,* $\mathcal{A}$*-search for* $\mathcal{E}$ *does not polynomial-time Turing reduce to* $\mathcal{A}$*-decision for* $\mathcal{E}$*.*

Let us present the idea behind the proof of Theorem 3.1, focusing in particular as an example on constructive control by run-off partition of candidates in the ties-promote model. So, let us use the statement's hypothesis, and assume that $\mathrm{P} \neq \mathrm{NP} \cap \mathrm{coNP}$ holds. We invoke a complexity-theoretic result known as the Borodin-Demers Theorem, which from that hypothesis yields the existence of an easily recognizable set of satisfiable formulas such that no polynomial-time machine can, for all of them, find a satisfying assignment. To the best of our knowledge, the Borodin-Demers Theorem has never before been applied in the study of elections, computational social choice, multiagent systems, or for that matter anywhere outside of computational complexity theory.

**Theorem 3.3 (The Borodin-Demers Theorem [BD76][8])** *If* $\mathrm{P} \neq \mathrm{NP} \cap \mathrm{coNP}$ *then there is a set* $B$ *so*

1. $B \in \mathrm{P}$,

[8]The provenance of this result is a bit tricky. The result is essentially due to Borodin and Demers [BD76] and to a result of Valiant [Val76] that they mention, which itself is focused on separating evaluating a function from checking a function. However, sharp-eyed readers will note that the actual theorem we state here cannot be found

2. $B \subseteq \mathrm{SAT}$, *and*

3. *no* P *machine can find solutions for all formulas in* $B$*. That is, for* no *polynomial-time computable function* $g$ *do we have* $(\forall f)[f \in B \Rightarrow g(f)$ *is a satisfying assignment of* $f]$.

So we have something quite striking: A set of boolean formulas that are easily recognized as being satisfiable but for which it is not in general easy to find how they can be satisfied, i.e., every polynomial-time machine fails on some of them (indeed, on infinitely many, as otherwise one could finitely patch). (Note: The Borodin-Demers Theorem does not establish, even conditionally, that search fails to reduce to decision for SAT; it is well-known that for SAT, search reduces to decision, unconditionally.) Our goal, of course, is to shoehorn the set $B$ into the world of election manipulation for a variety of manipulative actions. Of course, each manipulative action comes with its own form and definition, and so for many such shoehorning is essentially impossible—as we show in Section 3.2. But for others, we can do this, sometimes smoothly and sometimes through extreme, difficult contortions. The difficulty is that the structure of many electoral manipulations, and our goal to realize a separation with respect even to some election system with a polynomial-time winner problem, very much ties our hands. And in fact, even for our results here, the different manipulative actions have enormously differing proofs, as each proof must be tailored to the manipulative action.

Nonetheless, the general approach is clear and shared, although the implementations and constructions differ wildly. The general approach is given a set $B$ from the Borodin-Demers result, we must build an election system $\mathcal{E}$, whose winner problem is in P, such that for our manipulative action the decision problem is in P but the search problem is not polynomial-time computable. To do this, our election system $\mathcal{E}$ will clearly need to be very much attuned to $B$. It typically will be interpreting voters, candidates, collections of voters, and collections of candidates as variously trying to specify a Borodin-Demers "puzzle"—i.e., an obviously satisfiable formula (a string $x \in B$), and it also will interpret some similar things about its input as trying to propose solutions to that puzzle.

To really explain how this works in practice would require going through the actual proofs, which we do provide in the proofs section. But to give an idea of the flavor, let us speak here in a high-level, handwaving way about a specific example (that is neither our hardest nor our easiest case), namely constructive control by run-off partition of candidates in the ties-promote model. Our scheme here is to hope that our input consists of two almost-copies of a Borodin-Demers puzzle $x$, namely that part of our input is $x0$ and $x1$, $x \in B$. In particular, we'll hope that the lexicographically two smallest candidates have those strings as their names. Suppose that the obviously satisfiable formula $x$ (for concreteness of this informal discussion) is 1000 bits long and has 27 variables. Then we will hope to have exactly $2 \cdot 27 = 54$ other candidates, who will all form a lexicographically contiguous segment starting at, say $0^{5 \cdot 1000}$, i.e., the first of the 54 candidates is named $0^{5000}$, the second is named $0^{4999}1$, the third is named $0^{4998}10$, and so on. Now, we'll interpret these strings as 27 pairs—the first two, the next two, and so on. And we'll set up our election system so that it will try to ensure that exactly one of each pair goes on one side of the partition in any partition that will lead to victory of $x0$. The election system if it sees in its candidate set $x0$, $x \in B$, will compute the size and number of variables of $x$, will see if it has in either of those papers, although it certainly can be readily established from their claims and in spirit is essentially the same. The form stated here, which is an elegant form that we believe was first framed by Hartmanis who always simply attributed it to [BD76], to our knowledge first appears in print in one of Hartmanis's students' theses [Hem87], which itself attributes it to [BD76]. The result in that form can easily be established from [BD76] or [Val76], and for completeness we mention that a direct proof of the form we use is available in the course notes [Hem07], which themselves refer to this simply and appropriately as the Borodin-Demers Theorem. This form has appeared in various other works, such as [HRW97, Rot99], and in its so-called unambiguous analogue, the paper [HH90].

the right collection of other candidates to indicate it has precisely one from each of the $\#\text{vars}(x)$ pairs, will then interpret the low-order bit of each of those pair-choices as the $i$th bit of a guessed satisfying assignment for $x$, and if that assignment does satisfy $x$, will make $x0$ the one and only winner. Also, the election system when its input contains $x1$, $x \in B$, will check that it has one candidate from each of the $\#\text{vars}(x)$ pairs (and nothing else), and if so $x1$ and only $x1$ will win—it does not in this case do any satisfiability check. A third and final case in which we will have a winner is if the candidate set is $\{x0, x1\}$, $x \in B$, in which case $x0$ and only $x0$ will win. And these three cases are the only ways to win.[9]

Now, recall that run-off partition splits the candidates into two groups for primary elections and then runs the winners of those against each other. If the input set is of just the dream-case form we have described, and we ask whether $x0$ can by run-off partition, ties-promote, be made a winner of the overall election, the answer is obviously "Yes," as $x \in B$ is satisfiable and so the partition that puts into one side of the partition $x0$ and precisely a set of one-per-pair candidates encoding a satisfying assignment and puts the rest on the other side will have $x0$ win its first-round contest, will have $x1$ win its first-round contest, and will have $x0$ win the second-round contest between $x0$ and $x1$.

But it is possible to see that if we have a polynomial-time algorithm for the search problem of how to make $x0$ win, that on the special input we just described, any search-problem output, i.e., any successful partition, will immediately make clear a satisfying assignment of $x$, as the election system in fact will force that. So if we had a search-problem polynomial-time algorithm, the third property (the one about no FP function always yielding solutions) of the Borodin-Demers set $B$ would be violated. So search for our election system is not polynomial-time computable.

But our election system clearly does have a P winner problem—it is just three simple cases to check.

So all that remains is to show that the decision problem for this control type is in P. Note that we need a P algorithm that works for *all* inputs—not just inputs so nice as to have our dream-case format. However, when one carefully checks everything, with the system very clearly specified, one can see that this holds also. This is the part that causes a large part of the complexity of the election system; for example, the simpler system without the $x1$ requirement will fail this requirement.

*This completes our informal proof sketch for that case.*

That control type was not the hardest we face. Run-off partition is symmetric, so we could in effect use the two first-round elections to ensure that the input candidate set was rich enough to allow the partition to "guess" every possible assignment for an input puzzle—or to when that does not hold ensure that no candidate wins the overall election. But in contrast partition by candidates is asymmetric, with some candidates getting a "bye" past the first round. To handle that case—keeping in mind that the election system sees only its input but won't in general know whether it is being invoked in a first or a second round—requires a far more complicated approach. As always, a key difficulty is to ensure that unanticipated inputs don't cause our manipulation-action's decision problem to fail to be in P—something that could happen if a given input drove that problem to in effect be asking: Does this obviously satisfiable $n$-variable formula have a satisfying assignment *that falls in some collection, to which the input limits us, of just a certain less-than-$2^n$-sized family of the possible assignments*?

Now, Theorem 3.1 gives twelve cases where $\mathrm{P} \neq \mathrm{NP} \cap \mathrm{coNP}$ implies the existence of a P-winner problem election system where for a particular manipulative action decision is easy but search is hard. It is natural to wonder whether the converses of some or all of these twelve results hold. We note that either all of the converses hold or none do, and which of those cases holds is identical to a long-open issue in complexity theory, namely, whether the converse of the Borodin-Demers Theorem holds. (What is known regarding the converse of the Borodin-Demers Theorem is that there is a relativized world in which the converse does not hold [IN88]—which does not speak to what holds in the real world—and for the unambiguous analogue of the Borodin-Demers Theorem the converse does hold [HH88].) In particular, let us call the right-hand side of the Borodin-Demers Theorem the "Borodin-Demers Condition."

[9]The voters will be ignored in this election system. Since partition elections always have their votes masked down to just the parts of the vote involving the candidates in that part of the partition, any coding of information into votes could be wildly distorted by the partitioning. And so for this system we set ourselves the challenge of achieving our goal purely through codings involving the candidates—and we show that this will work.

**Definition 3.4 (The Borodin-Demers Condition)** *We say the Borodin-Demers Condition holds if there exists a set $B$ such that*

1. $B \in \mathrm{P}$,
2. $B \subseteq \mathrm{SAT}$, *and*
3. *no* P *machine can find solutions for all formulas in $B$. That is, for* no *polynomial-time computable function $g$ do we have $(\forall f)[f \in B \Rightarrow g(f)$ is a satisfying assignment of $f]$.*

In each of our twelve "$\mathrm{P} \neq \mathrm{NP} \cap \mathrm{coNP} \Rightarrow$" results about elections, we used "$\mathrm{P} \neq \mathrm{NP} \cap \mathrm{coNP}$" to note, by the Borodin-Demers Theorem, that the Borodin-Demers Condition held, and then we (in the proofs) do the often difficult work to show that the Borodin-Demers Condition implies that for the given type of manipulative action there exists an election system $\mathcal{E}$, with a polynomial-time winner problem, for which the decision version of that action is in P but the search version is not polynomial-time computable. The following result—which unlike some of the twelve cases of Theorem 3.1 is not hard to prove—shows that each of those twelve right-hand sides about elections implies the Borodin-Demers Condition, and thus that those twelve right-hand sides are all equivalent both to each other and to the Borodin-Demers condition. Also each of our twelve "$\mathrm{P} \neq \mathrm{NP} \cap \mathrm{coNP} \Rightarrow$" results about elections will have its converse hold if and only if the converse of the Borodin-Demers Theorem holds.

**Theorem 3.5** *For each of the twelve manipulative actions $\mathcal{A}$ listed in Theorem 3.1, the following two conditions are equivalent:*

- *The Borodin-Demers Condition holds.*
- *There exists an election system $\mathcal{E}$, with a polynomial-time winner problem, such that the $\mathcal{A}$ decision problem for $\mathcal{E}$ is in* P *but the $\mathcal{A}$ search problem for $\mathcal{E}$ is not polynomial-time computable.*

## 3.2 Cases Where Search Reduces to Decision

This section's central result states that for many types of manipulative actions search polynomial-time Turing reduces to decision.

**Theorem 3.6** *For each manipulative action $\mathcal{A}$ belonging to the following list:*

1. *constructive control by adding voters,*
2. *destructive control by adding voters,*
3. *constructive control by deleting voters,*

4. *destructive control by deleting voters,*

5. *constructive control by adding candidates,*

6. *destructive control by adding candidates,*

7. *constructive control by deleting candidates,*

8. *destructive control by deleting candidates,*

9. *destructive control by partition of candidates, ties promote,*

10. *destructive control by partition of candidates, ties eliminate,*

11. *destructive control by run-off partition of candidates, ties promote,*

12. *destructive control by run-off partition of candidates, ties eliminate,*

13. *constructive control by unlimited adding of candidates, and*

14. *destructive control by unlimited adding of candidates,*

*and for each election system $\mathcal{E}$, the $\mathcal{A}$ search problem for $\mathcal{E}$ polynomial-time Turing reduces to the $\mathcal{A}$ decision problem for $\mathcal{E}$.*

This theorem immediately implies that the behavior displayed in Theorem 3.1 is impossible for all of the above manipulative actions, even if Theorem 3.1's "winner problem in P" requirement is dropped.

**Corollary 3.7** *For each of the manipulative actions $\mathcal{A}$ listed in Theorem 3.6, for no election system $\mathcal{E}$ can it be the case that the $\mathcal{A}$ decision problem for $\mathcal{E}$ is in* P *but the $\mathcal{A}$ search problem for $\mathcal{E}$ is not polynomial-time computable.*

To make clear the flavor of the proofs of Theorem 3.6's parts, we will now prove four of those parts; the remainder will be proven in the proofs section, Section 4. We will in particular prove the result for the cases: destructive run-off partition of candidates, ties promote; destructive run-off partition of candidates, ties eliminate; destructive partition of candidates, ties promote; and destructive partition of candidates, ties eliminate. We first state the following result, which says that two pairs of these control types, which in all previous papers have been assumed to be distinct, are in fact identical. (We here use the same type of clear notational shorthand we used at the start of Section 3, as it stresses that we are claiming that sets are equal.)

**Theorem 3.8 (corollary to characterizations in a proof of [FHHR09a])**

1. DC-RPC-TP = DC-PC-TP *(i.e., viewed as decision problems, destructive control by run-off partition of candidates in the ties-promote model is* exactly *the same problem—the same set of strings—as is destructive control by partition of candidates in the ties-promote model).*

2. DC-RPC-TE = DC-PC-TE.

So that the present paper is complete and self-contained, we will prove Theorem 3.8 in Section 4. However, to support our proof of Theorem 3.6, all we now need is to briefly review the proof, as it is the proof rather than the theorem that we will need to draw on. The proof is simply that both DC-RPC-TP and DC-PC-TP in any election system $\mathcal{E}$ share the same characterization: In each case, $p$ can be prevented from winning by that type of control if and only if there exists a set $C'$, $\{p\} \subseteq C' \subseteq C$, such that in the $\mathcal{E}$ election $(C', V)$, $p$ is not a winner. Similarly, DC-RPC-TE and DC-PC-TE share the characterization that $p$ can be prevented from winning by that type of control if and only if there exists a set $C'$, $\{p\} \subseteq C' \subseteq C$, such that in the $\mathcal{E}$ election $(C', V)$ either there are no winners, two or more winners, or one winner with that winner not being $p$. Since these two pairs each share the same characterization, we have the two equalities. Curiously, the four characterizations in question all appear in passing (in the context of a proof about a particular election scheme "Copeland$^{\alpha}$," but the characterizations given apply in general) in a proof in [FHHR09a], in particular, that paper's argument that "Copeland$^{\alpha}$" is "vulnerable" to destructive control by RPC-TE, RPC-TP, PC-TE, and PC-TP. The only step that paper failed to take was to note that shared characterizations mean identical control types; due to the failure to take that small step, that paper and indeed all other classification papers dealing with these control types handle them as if they were distinct types, in some cases similar, separate proofs for types that in fact are outright identical.

The characterizations we just mentioned will be very useful in proving the DC-PC-TP, DC-PC-TE, DC-RPC-TP, and DC-RPC-TE cases of Theorem 3.6.

**Proof of Theorem 3.6, DC-PC-TP, DC-PC-TE, DC-RPC-TP, and DC-RPC-TE cases.** Let us show that for each election system $\mathcal{E}$ the search version of the DC-RPC-TP problem $\leq_T^p$-reduces to the decision version of the DC-RPC-TP problem. It is essential to keep in mind that we may *not* assume that the winner problem for $\mathcal{E}$ is in P. Our strategy will be to find, if any such exist, some *minimal* (not necessarily minimum-size, but rather minimal) set $C'$, $\{p\} \subseteq C' \subseteq C$, such that $p$ is not a winner of $(C', V)$. (By a minimal such set we mean here that for every set $C''$ satisfying $C'' \subsetneq C'$ and $p \in C''$ it holds that $p$ is a winner of the $\mathcal{E}$ election $(C'', V)$.)

So, suppose we are given some syntactically legal input (syntactically illegal input obviously have no successful action possible), consisting of $C$, $V$, and $p$. We cannot, for example, directly test whether the partition $(C, \emptyset)$ will kill off $p$ in the first round, as we may not assume that the $\mathcal{E}$ winner problem is in P. We must use our access to the decision problem to guide us. Recall that by the characterization mentioned before this proof, all we want to know is whether there exists a set $C'$, $\{p\} \subseteq C' \subseteq C$, such that $p$ is not a winner of the $\mathcal{E}$ election $(C', V)$. So ask $(C, V, p)$ to the decision version of DC-RPC-TP. If $(C, V, p)$ is not a member of that, no such $C'$ exists and so our search problem can state that no partition will succeed.

But if $(C, V, p)$ belongs to the decision problem, we know that such a $C'$ exists and merely need to find it. So, for each $c \in C$, $c \neq p$, in turn call the DC-RPC-TP decision problem on $(C - \{c\}, V, p)$. If all these calls are not in the decision version then $C$ is a set containing $p$ such that $p$ is not a winner of $(C, V)$—indeed it is a minimal such set. So return the partition $(C, \emptyset)$ as our search's result. If at least one of the $(C - \{c\}, V, p)$ was in the decision version, make $C - \{c\}$ our new "$C$," and repeat the above process iteratively, until we eventually get to (and recognize as such, as above) a minimal set $C'$ of the sort we are seeking. This set $C'$ shows that the partition $(C', C - C')$ will eliminate $p$ in the first round. And by the characterization, this algorithm will provide a good partition whenever destructive control is possible and in all other cases will detect that it is not.

For DC-PC-TP, search $\leq_T^p$-reduces to decision by essentially the same argument. One cannot directly conclude that from DC-RPC-TP = DC-PC-TP, as that equality is about the decision versions. But in light of the characterization of DC-PC-TP in terms of deleting candidates, all we need is to find a $C'$, $\{p\} \subseteq C' \subseteq C$, such that $p$ loses, and so we just like above can reduce DC-PC-TP search to its own decision version (which by the characterization is about deleting candidates, and that also happens to be the same set as DC-RPC-TP's decision version) and as above can find, when one exists, a minimal such set, $C'$. The only thing to be careful about is to make sure that, due to the asymmetry of PC, we make $C'$ the side of the partition that competes in the first round and make $C - C'$ the part of the partition that gets a first-round bye.

The proofs that the search version of the DC-RPC-TE problem $\leq_T^p$-reduces to its decision version, and that the search version of the DC-PC-TE problem $\leq_T^p$-reduces to its decision version are exactly analogous to the above proofs, altered to use the slightly different characterization throughout (including in the definition of the $C'$ we seek), i.e., seeking a minimal $C'$, $\{p\} \subseteq C' \subseteq C$, such that $p$ is not a unique winner of the election $(C', V)$. ❑

# 4 Proofs

In this section we provide all proofs and proof parts not provided earlier. For clarity, proofs—and in some cases proofs of parts of a theorem—are given in their own subsections, so that a few words can be said before each as to the proof—its intuition or its approach to coding or so on. This section does contain an additional result, namely Theorem 4.1.

## 4.1 Proofs for Section 3.1

We give the proofs for Section 3.1.

All of these proofs will have a similar structure. We start by assuming that $\mathrm{P} \neq \mathrm{NP} \cap \mathrm{coNP}$. Then we fix a set $B$ that fulfills the conditions of the Borodin-Demers Theorem, i.e., $B$ is a set that is in P, is a subset of SAT, and for which there is no polynomial-time function that for every formula in $B$ finds a satisfying assignment for that formula. We will then define a voting system using $B$, show that its winner problem is in P, show that the decision problem for the relevant manipulative action is in P, and then show that there can be no polynomial-time function for the search problem for the manipulative action. This last task will be accomplished through showing that we can in polynomial time encode a formula from our Borodin-Demers set $B$ into the search problem of the relevant manipulative action, in such a way that a solution to this search problem gives a satisfying assignment for the formula.

### 4.1.1 Theorem 3.1, Constructive Manipulation Case

This proof is likely the simplest of this subsection and closely follows the outline above. We construct an election system where there are no winners unless the candidates in the election specify a formula from our Borodin-Demers set $B$ and a voter specifies a satisfying assignment for this formula. This makes the decision version of the manipulation problem trivial, but it makes finding a successful manipulation equivalent to finding a satisfying assignment for a formula from $B$.

**Proof of Theorem 3.1, constructive manipulation case.** Assume that $\mathrm{P} \neq \mathrm{NP} \cap \mathrm{coNP}$ and let $B$ be a set that fulfills the conditions of the Borodin-Demers Theorem. We assume that our encoding of SAT is such that for each string $x \in \mathrm{SAT}$, the number of distinct variables in formula $x$ is at most $\max(0, \lfloor \frac{|x|}{2} \rfloor - 1)$.

In our voting system, the candidate names specify a string, puzzle($C$), as follows: Sort the candidate names lexicographically and then in order take the low-order bit from each name other than $\epsilon$ and concatenate those. Each voter will also specify a string, which we will use as an assignment to the variables of puzzle($C$), as follows. Recall that each vote is given as a tie-free linear ordering of the candidates. Sort the candidates of the election in lexicographic order. Each candidate with odd rank in this sorting will be assigned bit value 1, and each candidate with even rank will be assigned bit value 0. So for example if $C = \{$"Alice", "Bob", "Carol", "David"$\}$, "Alice" and "Carol" are 1's and "Bob" and "David" are 0's. Let $\mathrm{bit}_C(c)$ be the bit value of candidate $c$ among the set of candidates $C$. In a vote $c_1 < c_2 < \cdots < c_k$, the bitstring we associate with the vote is $\mathrm{bit}_C(c_1) \cdot \mathrm{bit}_C(c_2) \cdot \cdots \cdot \mathrm{bit}_C(c_{\lfloor \frac{k}{2} \rfloor - 1})$. Note that this allows us to have a voter put his or her favorite candidate as his or her top choice, regardless of the bitstring the voter is going to represent and whether that candidate is a 0 or a 1, since we have enough 0's and 1's to form any string in $\{0,1\}^{\max(0, \lfloor \frac{k}{2} \rfloor - 1)}$ and place any candidate out of the encoding region in first place.

Now we will specify the actual election system, which we will call $\mathcal{E}_1$. On any input $(C, V)$ we first compute puzzle($C$). If puzzle($C$) $\notin B$, then every candidate loses. If puzzle($C$) $\in B$, we examine every voter $v_i \in V$ and compute the bitstring $b_i$ of that voter as specified. Note that this bitstring has $\max(0, \lfloor \frac{||C||}{2} \rfloor - 1)$ bits. By our assumption about the encoding of SAT (and noting that $|\mathrm{puzzle}(C)| \leq ||C||$), this will give us enough bits to specify an assignment to the variables of puzzle($C$). If the first $d$ bits (where puzzle($C$) has $d$ distinct variables) of $b_i$ form a satisfying assignment for puzzle($C$), then the top choice of $v_i$ will be a winner.[10] We do this for all $v_i$, so the complete winner set is the set of all candidates who are the top choice of at least one voter that has a solution to the puzzle. This completes our specification of $\mathcal{E}_1$.

Now we have several things to show about this voting system to complete our proof.

First of all, the winner problem for $\mathcal{E}_1$ is in P. Why? $B \in \mathrm{P}$ and puzzle($C$) is easy to build from $C$, so detecting whether puzzle($C$) is in $B$ is easy. In the case that puzzle($C$) $\notin B$, no one wins, and we are done. In the case that puzzle($C$) $\in B$, we have to compute and check the solution attempt of each voter. However, this is easy as well since computing each $b_i$ is easy and testing whether an assignment satisfies a formula can easily be done in polynomial time (we can even do it in LOGSPACE, and with the right encoding even in ALOGTIME [Bus87]). Then we just need to toss the top choice of each successful vote into the winner set and we are done.

Next, the manipulation decision problem for $\mathcal{E}_1$ is in P. Why? If puzzle($C$) $\notin B$, then no one ever wins so manipulation cannot succeed. If puzzle($C$) $\in B$ and there is at least one manipulator whose vote we can change, there will exist a successful manipulation to make any candidate $p$ a winner. Namely, set that manipulator to a vote that gives a correct answer to the puzzle in its lower-order bits and has $p$ as its top choice. If puzzle($C$) $\in B$ but we cannot change any votes, the only possible manipulation is the empty one. Thus $p$ can be made a winner if and only if it is already a winner of the election, and we can determine this as the winner problem is in P. So the problem of determining whether manipulation is possible is easy.

Finally, the manipulation search problem for $\mathcal{E}_1$ is not in FP. Why? If it were in FP $B$ would violate its conditions. In particular, given a string $F \in B$, we could build a candidate set $\widehat{C}$ encoding $F$ as its puzzle. We will have just one voter, who will be manipulative. Let $p$ be any arbitrary candidate in $C$. An algorithm for the manipulation search problem would give us a vote that specifies a satisfying assignment for $F$. Thus we could use a polynomial-time function for the manipulation search problem to build a polynomial-time function that for every formula in $B$ finds a satisfying assignment for that formula. Such a function cannot exist by the third item of the

[10] We could also have defined every candidate to be a winner in this case, though this gives a less natural election system.

Borodin-Demers Theorem, so the constructive manipulation search problem must not be in FP for $\mathcal{E}_1$. ❑

#### 4.1.2 Theorem 3.1, Destructive Manipulation Case

This case will heavily draw from our previous proof for the constructive manipulation case, but we will need to construct a slightly altered voting system such that voters must "solve the puzzle" in order to prevent a candidate from winning, rather than to cause a candidate to win.

**Proof of Theorem 3.1, destructive manipulation case.** Again we assume that $P \neq NP \cap coNP$, we let $B$ be a set that fulfills the conditions of the Borodin-Demers Theorem, and we assume that our encoding of SAT is such that for each string $x \in SAT$, the number of distinct variables in formula $x$ is at most $\max(0, \lfloor \frac{|x|}{2} \rfloor - 1)$. We will now describe the voting system $\mathcal{E}_2$ that will show the separation of search and decision for destructive manipulation.

In $\mathcal{E}_2$ we will interpret the set of candidates as specifying a "puzzle" in the same way as with $\mathcal{E}_1$, and we will use the same technique for extracting a bitstring from each vote as well. $\mathcal{E}_2$ will then operate as follows. First we must compute puzzle($C$) from the candidate set $C$. If puzzle($C$) $\notin B$, then no candidate wins. In the case that puzzle($C$) $\in B$, we extract the bitstring $b_i$ from each voter $v_i$ and check if any of them give a satisfying assignment for puzzle($C$). If so, then there are no winners, but if not, every candidate is a winner. This ends our specification of $\mathcal{E}_2$.

Again we have several things to show about $\mathcal{E}_2$. First, we show that the winner problem is in P. As before, computing the puzzle and checking for membership in $B$ are both in P, so in the case that puzzle($C$) $\notin B$ we find the (empty) set of winners easily. If puzzle($C$) $\in B$, we also have to extract the bitstring from each vote and check particular assignments to puzzle($C$), but again this is easily doable in polynomial time. Thus the winner problem is easily in P.

We also need that the destructive manipulation decision problem is in P. In any case where puzzle($C$) $\notin B$, there are no winners, so we have a positive instance. If puzzle($C$) $\in B$ and the instance allows us to change the vote of a manipulator, then we have a positive instance, as setting the manipulator vote to specify a satisfying assignment for puzzle($C$) will cause no candidate to win the election, including the hated candidate. If we do not have the ability to assign a manipulator vote, we can determine if $p$ will win using the winner problem for $\mathcal{E}_2$, which we already saw was in P. We can easily perform any of these steps, including computing the puzzle, checking for membership in $B$, and checking assignments to a boolean formula, so the destructive manipulation decision problem is in P for $\mathcal{E}_2$.

Finally we need that the destructive manipulation search problem is not in FP for $\mathcal{E}_2$. We will again show that the existence of a polynomial-time search algorithm for destructive manipulation leads to the existence of a polynomial-time algorithm that for every formula in $B$ finds a satisfying assignment for that formula. This can proceed exactly as with constructive manipulation. Let $F$ be a string in $B$. We will then build a candidate set $\widehat{C}$ encoding $F$, we will have just one voter, who will be manipulative, and we let $p$ be any arbitrary candidate in $\widehat{C}$. Finding a successful manipulation to make $p$ not win will give us a satisfying assignment for $F$. So we have a polynomial-time function that for every formula in $B$ finds a satisfying assignment for that formula. But this cannot exist by the third item of the Borodin-Demers Theorem, so the destructive manipulation search problem must not be in FP for $\mathcal{E}_2$. ❑

#### 4.1.3 Theorem 3.1, Constructive Bribery Case

For this case we can reuse the voting system $\mathcal{E}_1$ from the constructive manipulation case and only slightly change the reduction from the Borodin-Demers set search problem so as to correspond with bribery rather than manipulation. Since we are using the same voting system we do not have to again show that the winner problem is in P but we will pick up from there.

**Proof of Theorem 3.1, constructive bribery case.** We are dealing with a different manipulative action, however, bribery instead of manipulation, so we will have to show that the constructive bribery decision problem is in P. We can easily compute puzzle($C$) and check for membership in $B$, and if puzzle($C$) $\notin B$, then there can be no winners and bribery will not succeed. On the other hand, if puzzle($C$) $\in B$, if there is at least one vote, and if we can bribe at least one voter, then this will be a positive instance, as bribing that voter to give a vote encoding a satisfying assignment together with the preferred candidate $p$ at the top of the voter's list will cause $p$ to be a winner. If there are not any voters or if we cannot bribe any voters, then we cannot alter the election, so we can check whether this is a positive instance simply with the winner problem which we have already seen is in P.

Now we will show that a polynomial-time function for the search version of this problem would lead to a polynomial-time function for finding satisfying assignments for the elements of $B$. Let $F$ be an element of $B$. We will build a candidate set $\widehat{C}$ encoding $F$ as its puzzle. We will then let the voter set $V$ contain a single voter with an arbitrary initial preference over the candidates and allow the bribery of a single voter. We run our hypothetical polynomial-time algorithm for the bribery search problem on this instance to find a vote that makes some candidate $p \in \widehat{C}$ win (if no bribery is needed, the initial preference will work). This vote must specify a satisfying assignment for $F$, and so we have a polynomial-time algorithm that for every formula in $B$ finds a satisfying assignment for that formula. Such a function cannot exist by the third item of the Borodin-Demers Condition, so the constructive bribery search problem must not be in FP for $\mathcal{E}_1$. ❑

#### 4.1.4 Theorem 3.1, Destructive Bribery Case

Here we can easily adapt our proof for destructive manipulation and use the same voting system $\mathcal{E}_2$ but just change the reduction from the Borodin-Demers set search problem to correspond with bribery, as in the previous proof. Again we have already shown that the winner problem for $\mathcal{E}_2$ is in P.

**Proof of Theorem 3.1, destructive bribery case.** We will show now that the decision problem destructive bribery is in P. As before we can easily compute the puzzle from the candidate set and test membership in $B$. If puzzle($C$) $\notin B$, there can be no winners so the destructive bribery instance will always succeed. If puzzle($C$) $\in B$, there is at least one voter, and we can bribe at least one voter, then we could make the hated candidate lose by changing a vote to specify a satisfying assignment for $B$. If there are not any voters or if we cannot bribe any voters, then we can not alter the election, so we can check whether this is a positive instance simply with the winner problem which we have already seen is in P.

We will now show that there cannot be a polynomial-time function for the destructive bribery search problem. This can proceed just as the constructive bribery case. Let $F$ be an element of $B$. We will build a candidate set $\widehat{C}$ encoding $F$ as its puzzle, include a single voter with an arbitrary initial preference over the candidates, and allow bribery of a single voter. The hated candidate $p$ can be any arbitrary candidate in the election. By running a hypothetical polynomial-time function for the destructive bribery search problem, we would acquire a changed vote for the single voter that would specify a satisfying assignment for $F$ (if no bribery is needed, the initial preference will work) and cause the election to have no winners. Therefore this could also be used to develop a polynomial-time algorithm that for every formula in $B$ finds a satisfying assignment for that formula. By the third item in the Borodin-Demers Condition, this cannot exist, so the destructive bribery search problem must not be in FP for $\mathcal{E}_2$. ❑

### 4.1.5 Theorem 3.1, Constructive Control by Partition of Voters (CC-PV) Case

Now we move into the control by partition cases, which are arguably harder and have a much different feel compared to the previous cases of manipulation and bribery. Now we need to build a voting system where the partition of the voters into two separate first-round subelections corresponds to an attempted solution to the "puzzle" of a Borodin-Demers formula. As in the previous cases, the candidate names specify the puzzle. Unlike the previous cases, we will use the set of voters to specify the assignment.

**Proof of Theorem 3.1, constructive control by partition of voters case.** We assume that $\mathrm{P} \neq \mathrm{NP} \cap \mathrm{coNP}$, we let $B$ be a set that fulfills the conditions of the Borodin-Demers Theorem, and we assume without loss of generality that every element of $B$ contains at least two distinct variables. We assume that our encoding of SAT is such that for each string $x \in \mathrm{SAT}$, the number of distinct variables in $x$ is at most $\lfloor \frac{|x|}{2} \rfloor$.

For this case, we use the same scheme as in $\mathcal{E}_1$ for computing puzzle($C$) from the candidate set. Suppose puzzle($C$) is a boolean formula with $d$ variables. Let $c_1, c_2, \ldots, c_{2d}$ be the first $2d$ candidates in the lexicographic order. This many candidates exist by our assumption on the encoding of SAT. Let $V$ be a set of voters. Consider the set of candidates that are the top choice of some voter in $V$. If that set is of the form $\{c_{1+\alpha_1}, c_{3+\alpha_2} \ldots, c_{2d-1+\alpha_d}\}$ with each $\alpha_i \in \{0,1\}$, we will interpret $\alpha_1 \cdot \alpha_2 \cdot \cdots \cdot \alpha_d$ as an assignment to the variables of puzzle($C$).

We will now describe our election system $\mathcal{E}_3$. Given an election $(C, V)$:

- If $||C|| \leq 2$, the lexicographically last candidate is the only winner.
- Else if $V = \emptyset$ or there exist two distinct voters in $V$ that have the same candidate as their top choice, the lexicographically first candidate is the only winner.
- Else if puzzle($C$) $\in B$ and $V$ specifies an assignment for puzzle($C$) as described above and this assignment is either a satisfying assignment or the complement of a satisfying assignment (that is, it is a satisfying assignment if we flip each 0 to a 1 and each 1 to a 0) for puzzle($C$), then the lexicographically first candidate is the only winner.
- Else the lexicographically last candidate is the only winner.

This completes our description of $\mathcal{E}_3$.

Since this voting system always has exactly one winner in a given election, it will have the exact same behavior for control by partition of voters in the ties promote and in the ties eliminate models. Thus we can prove our desired result for both of these cases simultaneously.

Now we again have several things to show about $\mathcal{E}_3$. First we need to show that its winner problem is in P. This is clear from the description of the voting system. Only in the third case are more than trivial checks performed, and still we must only compute the puzzle from the candidates and the variable assignment from the voters, check for membership in $B$, and check at most two assignments for a boolean formula, all of which can easily be performed in polynomial time.

We next need to show that the CC-PV decision problem is in P for $\mathcal{E}_3$. If there are at most two candidates, the lexicographically last candidate will be the only winner of both subelections (regardless of the voter partition) and then will be the only winner in the final election. So control in this case is possible if and only if the designated candidate is the lexicographically last. We claim that constructive control by partition of voters on input $((C, V), p)$ for $\|C\| > 2$ is possible if and only if:

- $p$ is the lexicographically last candidate and $V \neq \emptyset$, or
- $p$ is the lexicographically first candidate and we are in one of the following, easily identifiable, four cases:
  - $V = \emptyset$, or
  - there exist two distinct voters in $V$ that have the same candidate as their top choice, or
  - puzzle($C$) $\in B$ and $V$ specifies a satisfying assignment for puzzle($C$) or the complement of a satisfying assignment for puzzle($C$), or
  - puzzle($C$) $\in B$ and the set of candidates that are the top choice of some voter in $V$ is the set of the first $2d$ candidates in the lexicographic order, where $d$ is the number of variables in puzzle($C$).

It is immediate that control is only possible if $p$ is the lexicographically first or the lexicographically last candidate. First assume that $p$ is the lexicographically last candidate. If $V = \emptyset$, the lexicographically first candidate will always be the only winner, and control is not possible. If $V \neq \emptyset$, then let $v$ be some element from $V$. The lexicographically last candidate is the only winner of $(C, \{v\})$, since if puzzle($C$) $\in B$, puzzle($C$) has at least two variables, and so $v$ will not specify an assignment for puzzle($C$). It follows that partition $(\{v\}, V - \{v\})$ makes $p$ the winner.

For the remainder of the proof, assume that $p$ is the lexicographically first candidate. We will first show that control is possible in all four cases listed above. If $V = \emptyset$, the lexicographically first candidate will always be the only winner. If there exist two distinct voters in $V$ that have the same candidate as their top choice, then the partition $(V, \emptyset)$ makes $p$ the winner. If puzzle($C$) $\in B$ and $V$ specifies a satisfying assignment for puzzle($C$) or the complement of a satisfying assignment for puzzle($C$), then the partition $(V, \emptyset)$ makes $p$ the winner. If puzzle($C$) $\in B$ and the set of candidates that are the top choice of some voter in $V$ is the set of the first $2d$ candidates in the lexicographic order, where $d$ is the number of variables in puzzle($C$), then we can partition $V$ into $V_1$ and $V_2$ such that $V_1$ specifies a satisfying assignment for puzzle($C$) and $V_2$ specifies the complement of this satisfying assignment. $p$ will be the only winner of both subelections, and wins the final election.

It remains to show that if we are not in one of the four cases listed above, control is not possible. Suppose for a contradiction that control is possible and that this control is witnessed by partition $(V_1, V_2)$. Because of the definition of election system $\mathcal{E}_3$, this implies that $p$ is the winner of $(C, V_1)$ and the winner of $(C, V_2)$. It follows that puzzle($C$) $\in B$ and that $V_1$ and $V_2$ specify a satisfying assignment or the complement of a satisfying assignment for puzzle($C$). Since the top choice of each voter in $V$ is different, it follows that $V_1$ and $V_2$ specify complementary assignments for puzzle($C$). This implies that the set of candidates that are the top choice of some voter in $V$ is the set of the first $2d$ candidates in the lexicographic order, where $d$ is the number of variables in puzzle($C$), which contradicts our assumption that we are not in one of the four cases listed above.

Finally we need to show that the CC-PV search problem is not in FP. Let $F$ be a formula from $B$. We will construct a CC-PV instance that will allow us to find a satisfying assignment for $F$ in polynomial time using a polynomial-time algorithm for the CC-PV search problem. We will build a candidate set $\widehat{C}$ encoding $F$ as its puzzle. The voter set will consist of $2d$ voters, where $d \geq 2$ is the number of distinct variables in $F$, and the top choice of the $i$th voter will be the lexicographically $i$th candidate. The preferred candidate $p$ will be the lexicographically first candidate. Now, we can query this instance to a hypothetical algorithm for the CC-PV search problem. We know it will succeed, as this is a positive instance of CC-PV, as explained above. Furthermore, the partition that the algorithm finds will necessarily give a satisfying assignment for $F$. We will have to arbitrarily pick one side of the partition and check whether the specified assignment or its complement is a satisfying assignment, but this can be done easily in polynomial time. Thus the existence of a polynomial-time algorithm for the CC-PV search problem leads to the existence of a polynomial-time algorithm that for every formula in $B$ finds a satisfying assignment for that formula. Such an algorithm cannot exist, so the CC-PV search problem must not be in FP for $\mathcal{E}_3$. ❑

#### 4.1.6 Theorem 3.1, Destructive Control by Partition of Voters (DC-PV) Case

In this case we can reuse the voting system $\mathcal{E}_3$ and many of the arguments from the constructive case.

**Proof of Theorem 3.1, destructive control by partition of voters case.** We have already shown that the winner problem is in P. To show that the DC-PV decision problem is in P for $\mathcal{E}_3$, it suffices to observe that since every $\mathcal{E}_3$ election has exactly one winner, destructive control by partition of voters on input $((C, V), p)$ is possible if and only if there exists a candidate $q \in C - \{p\}$ such that constructive control by partition of voters on input $((C, V), p)$ is possible. Finally, to show that the DC-PV search problem is not in FP, we encode $F \in B$ in the same way as in the constructive case. Let $p$ be the lexicographically last candidate. As in the constructive case, a successful control action is possible, and we can easily compute a satisfying assignment for $F$ from a partition witnessing this success. ❑

#### 4.1.7 Theorem 3.1, Constructive Control by Partition of Candidates, Ties Promote (CC-PC-TP) Case

The control by partition of candidates cases of control have a strange asymmetry unlike the partition of voters cases and the run-off partition of candidates cases. This makes this proof and the next perhaps the trickiest we have here. The basic idea is the same as the others in this section, that we design an election system that encodes a Borodin-Demers puzzle in a way that we must (in some limited cases at least) partition according to a satisfying assignment in order to succeed at control. There are however a number of differences here, as now we will use the candidate names to specify both the puzzle and the solution. And in some cases we will have many winners in a given election, unlike our other partition cases that only ever had one. This limits this proof to only work for the TP model but allows us to deal with the asymmetry of the control action.

**Proof of Theorem 3.1, constructive control by partition of candidates, ties promote case.** We assume that $\mathrm{P} \neq \mathrm{NP} \cap \mathrm{coNP}$, we let $B$ be a set that fulfills the conditions of the Borodin-Demers Theorem, and we assume that our encoding of SAT is such that for each string $x \in \mathrm{SAT}$, the number of distinct variables in formula $x$ is at most $|x|$. We will now describe the voting system which we will call $\mathcal{E}_4$. As we mentioned, we will now use candidate names to specify both the puzzle and the attempted solutions. Some candidates are of the form $0puz0$, where $puz$ is the Borodin-Demers puzzle. Other candidates are of the form $10puz010^{5k} + 2i + \alpha_i$, where $k$ is the length of $puz$ and "+" denotes the lexicographic increment. This candidate would be interpreted as specifying both the puzzle and the assignment of a value of $\alpha_i$ to the $i$th variable in $puz$. We will now specify the action of the voting system. If $C = \{0puz0\} \cup \bigcup_{1 \leq i \leq d} \{10puz010^{5k} + 2i + \alpha_i\}$, where $puz \in B$, $k$ is the length of $puz$, $d$ is the number of variables in $puz$, $\alpha_i \in \{0, 1\}$, and the full string $\alpha$ composed of all the bits $\alpha_1, \ldots, \alpha_d$ is either a satisfying assignment for $puz$ or the complement of one, then $0puz0$ will be the only winner. If $C$ consists of $0puz0$ along with any subset of $\bigcup_{1\leq i\leq d}\{10puz010^{5k}+2i, 10puz010^{5k}+2i+1\}$, where $puz \in B$, $k$ is the length of $puz$, $d$ is the number of variables in $puz$, $\alpha_i \in \{0,1\}$, and we are not in the previous case, then every candidate except $0puz0$ will be a winner. In every other case every candidate will be a winner. This completes our specification of $\mathcal{E}_4$. Note that unlike some of our other systems we totally ignore any and all of the voters in this system.

First, we must show that the winner problem for $\mathcal{E}_4$ is in P. This is easy, as though the specific cases are complicated in form they are simple to recognize. The additional checks of detecting whether $puz$ is in $B$ and checking whether $\alpha$ is a satisfying assignment for $puz$ or the complement of one can also be accomplished in polynomial time.

Next we must show that the CC-PC-TP decision problem is in P for $\mathcal{E}_4$. If $C \not\subseteq \{0puz0\} \cup \bigcup_{1\leq i\leq d}\{10puz010^{5k}+2i, 10puz010^{5k}+2i+1\}$, where $puz \in B$, $k$ is the length of $puz$, $d$ is the number of variables in $puz$, and $\alpha_i \in \{0,1\}$, then control will always be possible, since all candidates are winners if we put no candidates in the first-round election. So suppose that $C \subseteq \{0puz0\} \cup \bigcup_{1\leq i\leq d}\{10puz010^{5k}+2i, 10puz010^{5k}+2i+1\}$, where $puz \in B$, $k$ is the length of $puz$, $d$ is the number of variables in $puz$, and $\alpha_i \in \{0,1\}$. If $p$ is not $0puz0$, then control can easily succeed as we could just put all candidates in the first round, eliminating only $0puz0$ in the first round and letting every other candidate win in the final election. If our distinguished candidate is $0puz0$, control can only succeed in a few specific but easy to recognize situations. One case is if we have the complete set of variable assignment candidates along with $0puz0$. In this case we can make $0puz0$ win by putting it together with a satisfying assignment of variable assignment candidates in the first round. Then it alone will move on to the final round and face the complement of this satisfying assignment in the final election and come out a winner. The only other case in which $0puz0$ can be made a winner through this type of control is if the candidate set consists of $0puz0$ along with exactly a satisfying assignment (or the complement of one) of the variable assignment candidates. In this case we can just give every candidate a bye and $0puz0$ will win the final election. These cases are all easy to recognize so the decision problem is easily in P.

Finally we need to show that the CC-PC-TP search problem is not in FP. Let $puz$ be an element from $B$, let $k$ be the length of $puz$, and let $d$ be the number of variables in $puz$. We will then construct a voter set $C$ as $\{0puz0\}\cup\bigcup_{1\leq i\leq d}\{10puz010^{5k}+2i, 10puz010^{5k}+2i+1\}$. The voter set can remain empty as this voting system totally ignores the voters. Our distinguished candidate $p$ will of course be $0puz0$. The only way to make $0puz0$ a winner through this type of control is to put it in the first-round election with variable assignment candidates specifying a satisfying assignment or the complement of a satisfying assignment. Thus a polynomial-time search function for CC-PC-TP will allow us to find a satisfying assignment for $puz$ and thus give us a polynomial-time algorithm that for every formula in $B$ finds a satisfying assignment. Such an algorithm cannot exist, so CC-PC-TP search must not be in FP for $\mathcal{E}_4$. ❑

#### 4.1.8 Theorem 3.1, Constructive Control by Partition of Candidates, Ties Eliminate (CC-PC-TE) Case

Unlike all our other proofs in this section, for partition of candidates we do need a separate proof for the ties-promote and ties-eliminate cases. The system $\mathcal{E}_4$ in our previous proof often selects large numbers of winners, so the behavior of that system will be different under the TP and TE models. In the system we use here, we ignore the votes completely, add several special candidates, and use candidates to specify both the puzzle and the assignment.

**Proof of Theorem 3.1, constructive control by partition of candidates, ties eliminate case.** We will use the variable assignment candidates from the previous proof to specify the puzzle and mark the variable assignments. We also have three special candidates named $a$, $b$, and $c$. For concreteness, let $a = 0$, $b = 01$, and $c = 10$. We assume that $\mathrm{P} \neq \mathrm{NP} \cap \mathrm{coNP}$, we let $B$ be a set that fulfills the conditions of the Borodin-Demers Theorem, we assume that our encoding of SAT is such that for each string $x \in \mathrm{SAT}$, the number of distinct variables in formula $x$ is at most $|x|$, and we assume that every $x \in B$ contains at least one variable. We will now describe the behavior of voting system $\mathcal{E}_5$ on input $(C, V)$.

- If $a \in C$ but $b, c \notin C$, winners $= \emptyset$.
- If $C = \{b\}$ or $C = \{c\}$, winners $= \emptyset$.
- If $b \in C$ but $a, c \notin C$ and at least one other candidate is in $C$, winners $= \{b\}$.
- If $c \in C$ but $a, b \notin C$ and at least one other candidate is in $C$, winners $= \{c\}$.
- If $b, c \in C$, winners $= \{b, c\}$.
- If $a, b \in C$ but $c \notin C$:

  If the other candidates in $C$ specify $puz \in B$ and a satisfying assignment for $puz$, then winners $= \{a\}$.

  If the other candidates in $C$ specify $puz \in B$ and a nonsatisfying assignment for $puz$, then winners $= \{a, b\}$.

  Else winners $= \emptyset$.
- If $a, c \in C$ but $b \notin C$:

  If the other candidates in $C$ specify $puz \in B$ and a satisfying assignment for $puz$, then winners $= \{a\}$.

  If the other candidates in $C$ specify $puz \in B$ and a nonsatisfying assignment for $puz$, then winners $= \{a, c\}$.

  Else winners $= \emptyset$.
- In every other case winners $= \emptyset$.

This system is more complicated than the others we have used so far and there are a few things to note. First of all, only $a$, $b$, and $c$ can ever win the election. No candidate wins an election where it is the only candidate. And $a$ can only ever win if it is together with exactly one of $b$ or $c$ and with a variable assignment. And $a$ can only be a unique winner if they are together with a satisfying variable assignment.

We will now show that the winner problem for $\mathcal{E}_5$ is in P. Although there are many separate cases specifying $\mathcal{E}_5$'s winner problem, most of them only require very simple checks to determine the presence of the special candidates. Only in the last two cases do we have to test of membership in our Borodin-Demers set $B$ or deal check for boolean formula satisfaction, but those things are easily in P as well. So the winner problem as whole is easily in P.

We will show next that the CC-PC-TE decision problem is in P for $\mathcal{E}_5$. Most of the work and special cases for this proof were directed towards making sure this would be the case. The main idea is that many instances of control are positive in this voting system, but they are very simple to detect (and for most of them it is simple to produce the exact control action as well). If the distinguished candidate is either $b$ or $c$ and there is at least one other candidate that is not $a$, then control will always be possible. The only circumstances that cause $b$ or $c$ to not be a winner is if they are alone in the election, or if they are in a particular case alongside $a$. As long as there is one other candidate, we can ensure $b$ or $c$ is a winner by giving every candidate except $a$ (if they are present) a bye. However if $a$ is a candidate, we can ensure it does not interfere with $b$ or $c$ becoming a winner by simply putting $a$ alone in the first-round subelection where it will be eliminated, thus protecting $b$ or $c$ in the final round. If the distinguished candidate is either $b$ or $c$ and there is not at least one other candidate that is not $a$, then control will never be possible.

If the distinguished candidate is $a$, we can succeed in a somewhat narrower set of circumstances. The perfect case is when $C$ consists of $\{a, b, c\}$ along with the complete set of variable assignment candidates (0 and 1 for every variable) for some $puz \in B$. In this case we put $a$, $b$, and a satisfying assignment for $puz$ in the first round. Candidate $a$ will thus be the only winner of the first-round election, and $a$ will also be a winner of the final election with $c$ and the complement of this satisfying assignment (which may or may not itself be a satisfying assignment). Another similar case is if $C$ consists of $\{a, b, c\}$ along with the a satisfying assignment for some $puz \in B$ and an assignment for a different $puz' \in B$. Another case where control can succeed for $a$ is if the candidate set contains $a$, and only one of $b$ or $c$, and $C$ contains an assignment for some $puz \in B$, by giving a bye to all candidates other than $a$, $b$, and $c$, and an assignment for $puz \in B$. Similarly we can make $a$ win if $C$ consists of $a$, $b$, and $c$ and an assignment for some $puz \in B$, by putting $c$ (or $b$) alone in the first round, thus eliminating it and ending up in a situation like the previous case. It is easy to check that these are the only cases in which control is possible. Though we have a large number of cases here, none of them are difficult to check and most simply require checking for the presence of the special candidates.

Finally we can show that CC-PC-TE search is not in FP. Let $puz$ be an element from $B$. We then construct an instance of CC-PC-TE search with an empty voter set, a candidate set consisting of $\{a, b, c\}$ together with a complete set of variable assignment candidates encoding $puz$ and having 0 and 1 assignments for every variable in $puz$, and let our preferred candidate be $a$. This follows the pattern of our "perfect" case mentioned earlier, so there must be a successful control action. The action in fact must be to partition either $a$ and $b$ or $a$ and $c$ along with assignment candidates specifying a satisfying assignment for $puz$ into the first-round election. Thus a polynomial-time algorithm for CC-PC-TE search would give us a polynomial-time algorithm for finding satisfying assignments for elements of $B$ as well. Such an algorithm cannot exist, so CC-PC-TE search must not be in FP for $\mathcal{E}_5$. ❑

#### 4.1.9 Theorem 3.1, Constructive Control by Run-Off Partition of Candidates (CC-RPC) Case

We already covered this case in some detail in Section 3.1 but we will cover it again here and fill in some missing details. This case is similar to the proofs for control by partition of voters except that since we are partitioning candidates we use the candidate names to specify both the puzzle and the assignment, and the voters are not even used.

**Proof of Theorem 3.1, constructive control by run-off partition of candidates case.** We again have schemes for interpreting some candidate names as the Borodin-Demers puzzle and some others as an assignment to the variables of this puzzle. We distinguish between the two by padding the assignment candidate names out to be much longer than the puzzle candidate names. The puzzle candidates will simply consist of the puzzle $x$ followed by a single bit, either a zero or a one, while the assignment candidates all will be five times the length of the puzzle candidates. Given that the puzzle $x$ is of length $k$, each of these will consist of a bitstring $0^{5k}$ incremented $2i$ places past to denote that the $i$th variable should have value 0, or incremented $2i+1$ places past to denote that the $i$th variable should have value 1.

We again of course assume that $P \neq NP \cap coNP$ and let $B$ be a Borodin-Demers set. Without loss of generality, we assume that every element of $B$ contains at least one variable and that $\epsilon \notin B$. We assume that our encoding of SAT is such that for each string $x \in$ SAT, the number of distinct variables in formula $x$ is at most $|x|$. We will now specify the behavior of the voting system $\mathcal{E}_5$ on an election $(C, V)$.

- If the lexicographically smallest candidate is $x0$ with $x \in B$, and the remainder of the candidate set consists of $d$ variable assignment candidates (where $x$ contains $d$ variables) giving exactly one value to each variable in $x$, and with the full set representing a satisfying assignment for $x$, then only $x0$ will win.
- If the lexicographically smallest candidate is $x1$ with $x \in B$, and the remainder of the candidate set consists of $d$ variable assignment candidates (where $x$ contains $d$ variables) giving exactly one value to each variable in $x$ (but with no requirement that we have a satisfying assignment), then only $x1$ will win.
- If the entire candidate set consists of two candidates $x0$ and $x1$ with $x \in B$, then only $x0$ wins the election.
- In every other case no one wins.

Note that in no case do we ever have more than a single winner, so the behavior of this system is exactly the same in the ties-promote and ties-eliminate models.

We can see that the winner problem for this voting system is easily in P. Only a few cases produce any winners at all and they are easy to check for. In the first we have to test whether the puzzle boolean formula is satisfied by a particular assignment but this is easily doable in polynomial time, and in the others we only have to test for the presence of certain candidates and test for membership in $B$, which is in P by definition.

The decision problem for this case of control is in P as well. First of all, no candidate other than $x0$ or $x1$ for $x \in B$ can ever win, and so any instance with any other preferred candidate will always be negative. Since we are dealing with RPC, every candidate must go through a first-round subelection and then win in the final election to win overall. And since we never have more than one winner, if a candidate is going to win overall it must win in a final election containing either one or two candidates. The only case where that may happen is in our third winner case above, where $x0$ wins among $x0$ and $x1$. So the only way to make any candidate win following RPC is to have $x0$ win one subelection, and $x1$ win the other, and to then have $x0$ win the final election. This can only happen if our entire candidate set consists of $x0$ (with $x \in B$), $x1$, and the entire set of variable assignment candidates for $x$, with candidates assigning 0 or 1 to every variable in $x$. And in this case $x0$ can win as described by being put together with a satisfying assignment of assignment candidates in the first round and with $x1$ being put together with the other candidates in the first round, leading to $\{x0, x1\}$ as the candidates in the final election. This situation is easily detectable so this form of control is in P.

Finally, the CC-RPC search problem is not in FP in either tie handling model. Let $x$ be an element from $B$ and let $d$ be the number of variables in $x$. We can then easily use a polynomial-time algorithm for the CC-RPC search problem to find a satisfying assignment for $x$. We can do this simply by building an election with $C = \{x0, x1\} \cup \bigcup_{1 \leq i \leq d} \{0^{5k} + 2i, 0^{5k} + 2i + 1\}$ and let $p = x0$. That is, we create the one election described above where CC-RPC has a chance to work. We then could use a polynomial-time search algorithm for this problem to find the successful partition, and this will yield a satisfying assignment for $x$. Thus a polynomial-time algorithm for the CC-RPC search problem would give us a polynomial-time algorithm that for every formula in $B$ finds a satisfying assignment for that formula. Such an algorithm cannot exist, so the CC-RPC search problem in either tie model must not be in FP for this voting system. ❑

## 4.2 Proofs for Section 3.2

We give the proofs for Section 3.2 that were not covered there. The proofs of the remaining parts of Theorem 3.6 are all quite similar. For each of the relevant manipulative actions we provide an algorithm for the search version of the manipulative action that is allowed to make use of a unit-cost subroutine for the decision version of the manipulative action. All proofs will use the self-reducibility of the decision version to obtain a solution to the search problem. As is typical, votes are viewed as a multiset but come in as a list of ballots, one per voter. However, we note that if votes are represented succinctly (i.e., as a set of pairs consisting of a preference and the frequency of that preference), search also reduces to decision for all the cases of Theorem 3.6. The candidate cases follow with exactly the same proof as for the nonsuccinct cases. The proofs for the voter cases are quite similar to the proofs for the nonsuccinct cases below, except that we now use the decision problem to find out, with binary search, how many voters with a particular preference will be added to/deleted from $V$ to get the successful control action.

### 4.2.1 Theorem 3.6, Control by Adding Voters Cases

We will show that the search versions of constructive and destructive control by adding voters polynomial-time Turing reduce to the decision versions by providing appropriate reductions. We will first consider the constructive case, as the destructive version of the algorithm can be easily adapted from it. The algorithm will proceed iteratively, making use of the self-reducibility of this problem. That is, an instance of control by adding voters can be successful if and only if it can be successful either with a particular voter added, or without that voter added.

**Proof of Theorem 3.6, control by adding voters cases.** The algorithm will proceed as follows. The input will be a candidate set $C$, sets of voters $V$ and $W$, a distinguished candidate $p \in C$, and an adding limit $K$ (a nonnegative integer).

We will first query the input instance to our CC-AV (constructive control by adding voters) decision problem subroutine. If the answer is "No," we will return a fixed value denoting failure. Otherwise, we know a successful control action exists and we will proceed to find it.

We will initialize our added voter set $S$ to $\emptyset$, and then proceed by iterating as long as $W$ is nonempty and $K > 0$. Let $w$ be some arbitrary voter from $W$. We will make a call to the decision problem for the instance $(C, V \cup \{w\}, W - \{w\}, p, K - 1)$. That is, we are asking if we can make $p$ a winner while adding $w$ and no more than $K - 1$ other voters. If so, we remove $w$ from $W$, add $w$ to both $V$ and to $S$, subtract one from $K$, and move on to the next iteration. If not, we instead just remove $w$ from $W$ and move on to the next iteration.

If we do not have a failed instance (and note that failure will not occur if it does not occur upon the initial query) we proceed until we either have exhausted all our voters from $W$ or the parameter $K$ is equal to zero. Note that if we have a positive instance of this case of control (in the constructive case) with either an empty $W$ or with $K = 0$, then $p$ must be a winner of the election $(C, V)$, as there is no more possibility to add voters. We can see that we can easily reduce from the winner problem for an election system to the decision version of this problem.

So at this point, our set $S$ will contain all the added voters and $p$ will win the election $(C, V \cup S)$ (using the original value for $V$). Thus $S$ is a set of voters such that $||S|| \leq K$ and $p$ is a winner of $(C, V \cup S)$. Thus we can return $S$ as our successful search action.

This algorithm, making use of a unit-cost subroutine for the decision version of CC-AV, will easily run in polynomial time, will output a successful manipulative action when one exists, and will otherwise signify failure. Thus this algorithm provides a polynomial-time Turing reduction from the decision version of CC-AV to the search version of that problem, for any voting system.

The algorithm for the constructive case can be reused almost exactly for the destructive case, with the single change that we must make a call to a subroutine for the destructive version of this problem rather than the constructive version. ❑

### 4.2.2 Theorem 3.6, Control by Deleting Voters Cases

For these cases, again we will exploit a sort of self-reducibility of the control by deleting voters problem to build a simple iterative algorithm that uses the decision version of the problem to extract a successful control action if one exists. In particular, an instance of control by deleting voters can be successful if and only if it can be successful either by deleting a particular voter or by not deleting that particular voter. As before, we will handle the constructive case first and then describe how to adapt the reduction we give to the destructive case.

**Proof of Theorem 3.6, control by deleting voters cases.** We will now give the reduction. The input to the algorithm will be an election $(C, V)$, a distinguished candidate $p \in C$, and a nonnegative integer $K$. We will first query the input instance to our CC-DV (constructive control by deleting voters) decision problem subroutine. If the answer is "No," we will return a fixed value denoting failure. Otherwise, we know a successful control action exists and we will continue.

We will initialize a set $S$ to $\emptyset$ in order to keep track of the set of voters to delete. We will then proceed to iterate over the set of voters $V$ and build our set of deleted voters. For each voter $v \in V$, and as long as $K > 0$, we query to our CC-DV decision problem the instance $((C, V - \{v\}), p, K - 1)$. That is, we ask if we can make $p$ a is winner by deleting $v$ and up to $K - 1$ other voters. If this is the case, we remove $v$ from $V$, add it to $S$, subtract one from $K$, and move on to the next iteration. If not, there must instead be a successful control action without $v$ deleted, so we just move on to the next iteration.

We proceed in this way until we go through the entire voter set, or until $K$ is zero. If we did not return failure at the start of the algorithm, we must then produce an $S$ such that $||S|| \leq K$ and $p$ is a winner of $(C, V - S)$ (with our original value for $V$). Thus we return a successful control action if one exists, and this algorithm will clearly run in polynomial time. Thus we have shown that search polynomial-time Turing reduces to decision for CC-DV.

As with the adding voters cases, we can provide an algorithm for the destructive version of this case almost exactly as for the constructive version of this control problem with the single change that we call a subroutine for the destructive version of the problem instead. ❑

### 4.2.3 Theorem 3.6, Control by Adding Candidates Cases

This case is almost the exact mirror of the cases for control by adding voters and we base the algorithm on a very similar self-reduction, except that we are now adding to the candidate set instead of the voter set. Again we will consider the constructive version first.

**Proof of Theorem 3.6, control by adding candidates cases.** We will now give the reduction from the search version of CC-AC to the decision version of that problem. The input to our algorithm will be disjoint candidate sets $C$ and $A$, a voter set $V$ with preferences over $C \cup A$, a distinguished candidate $p \in C$, and a nonnegative integer $K$, our adding limit.

Initially we will query our input to our subroutine for the decision version of CC-AC (constructive control by adding candidates). If the answer is "No," we will return a fixed value denoting failure. Otherwise, we know a successful control action exists and we will go on to find it.

First we must initialize our added set $S$ to be $\emptyset$. We will then iterate as long as $A$ is not empty and $K > 0$. First let $a$ be some element of $A$. We will use the decision problem to determine whether we should add $a$ or not. We will query to the decision problem the instance $(C \cup \{a\}, A - \{a\}, V, p, K - 1)$. That is, we are asking if we can make $p$ a winner by adding $a$ and up to $K - 1$ other candidates. If this comes back positive, we add $a$ to $S$ and to $C$, and remove it from $A$. We also subtract one from $K$. If it comes back negative, we remove $a$ from $A$ and move on to the next iteration, as in this case it must be possible to make $p$ win while not adding $a$.

We continue until either $K = 0$ or $A = \emptyset$. At this point, there is no more allowance for adding candidates and changing the election, so if did not hit failure, $p$ will clearly win the election $(C \cup S, V)$ (with the original value for $C$). So we will return $S$, and $S$ will be a successful control action. In the cases where control is not possible we will indicate failure after the first query, and this algorithm clearly runs in polynomial time. Thus we have shown that search polynomial-time Turing reduces to decision for CC-AC for any voting system.

As with the previously handled voter control cases, we can exploit the same self-reducing structure as with the constructive version of this control problem and use essentially the same algorithm as for the constructive version of this problem but just use the appropriate destructive version for the subroutine call. ❑

#### 4.2.4 Theorem 3.6, Control by Deleting Candidates Cases

This case is closely parallel to the proof for control by deleting voters and we will use a similar self-reductive structure to build this algorithm, except of course modifying the candidate set throughout instead of the voter set. As before we will first deal with the constructive case.

**Proof of Theorem 3.6, control by deleting candidates cases.** The algorithm will proceed as follows. We will take as input an election $(C, V)$, a distinguished candidate $p \in C$, and a nonnegative integer $K$, our deletion limit. We will first query our subroutine for the decision version of CC-DC (constructive control by deleting candidates) with the input instance. If the answer is "No," we will return a fixed value denoting failure. Otherwise, we know a successful control action exists and we will proceed to find it.

We again initialize a set $S$ to $\emptyset$ to keep track of our deleted candidates. We will iterate as long as $K > 0$ through each $c \in C$, except for $p$, and do the following. We will query to the decision problem the instance $((C - \{c\}, V), p, K - 1)$. That is, we are asking if $p$ can be made a winner with $c$ deleted and with up to $K - 1$ other candidates from $C$ deleted. If so, we remove $c$ from $C$, add $c$ to $S$, and subtract one from $K$. If not, we know that there must be some successful control action without $c$ deleted, so we just move on to the next iteration.

We continue until either we have iterated through the entire set $C$ (except for $p$) or $K = 0$. At this point $p$ will be a winner of the election $(C - S, V)$ with $S$ such that $||S|| \leq K$ (and with our original $C$). We thus return $S$ as the successful control action. If there is no successful control action, we would have indicated failure after the initial query. Also, this algorithm clearly runs in polynomial time, so we have that search polynomial-time Turing reduces to decision for CC-DC for any voting system.

Again we can reuse this algorithm for the destructive version of this problem, only changing the subroutine call to use the destructive version of the decision problem. ❑

#### 4.2.5 Theorem 3.6, Control by Unlimited Adding of Candidates Cases

The case differs slightly from the others in that we do not have a parameter limiting the number of candidates that can be added, and so we will always have to iterate through the entire additional candidate set before generating the control action, but otherwise it follows essentially the same self-reducing structure as the previous proofs. Again, we will first handle the constructive case.

**Proof of Theorem 3.6, control by unlimited adding of candidates case.** Our algorithm will proceed as follows. We are given as input disjoint sets of candidates $C$ and $A$, a voter set $V$ with preferences over $C \cup A$, and a preferred candidate $p \in C$. We will first query the input instance to our CC-ACU (constructive control by unlimited adding of candidates) decision problem subroutine. If the answer is "No," we will return a fixed value denoting failure. Otherwise, we know a successful control action exists and we will continue.

As before we will use a set $S$, initialized to $\emptyset$, to keep track of the candidates we add. We iterate as long as $A$ is not empty. Let $a$ be some element of $A$. We query to the decision problem the instance $(C \cup \{a\}, A - \{a\}, V, p)$. That is, we are asking if we can make $p$ the winner by adding $a$ and possibly some other candidates from $A$ to the election. If so, we add $a$ to $C$ and to $S$, remove it from $A$, and go on to the next iteration. If not, we remove $a$ from $A$ and go on to the next iteration, as there must be some way to make $p$ win when not adding $a$.

We continue until $A = \emptyset$ and then return $S$. At this point, if we did not indicate failure (which again can only happen following the initial query), $p$ will be a winner of the election $(C \cup S, V)$ (using the original value for $C$) and $S$ will give us the successful control action we sought. Otherwise we quickly detect and indicate when control cannot succeed, and the whole algorithm clearly runs in polynomial time. Thus we have shown that search polynomial-time Turing reduces to decision for CC-ACU for any voting system.

Again, we only must change the subroutine call from the previous proof to use the destructive version the decision problem to obtain a polynomial-time function for the destructive version of this problem. ❑

#### 4.2.6 Theorem 3.8

We will now prove our earlier-used result that DC-RPC-TP and DC-PC-TP, and DC-RPC-TE and DC-RPC-TE are in fact pairwise identical problems, that is, they are the same sets. We do this by finding simple equivalent characterizations of these problems, and by finding identical characterizations for the both of the problems in both of those pairs.

**Proof of Theorem 3.8.** Let $\mathcal{E}$ be any election system. Consider any input to the DC-RPC-TP problem (by "input to the DC-RPC-TP problem" we mean any string in $\Sigma^\star$, not just those that belong to the set DC-RPC-TP; for syntactically illegal input we tacitly take it that we treat them as not belonging to the set DC-RPC-TP). (Keep also in mind that our entire paper is in the nonunique-winner model.) When can destructive control of candidate $p \in C$ succeed in the RPC-TP model? We claim destructive control can succeed if and only if there exists a set $C'$ such that $\{p\} \subseteq C' \subseteq C$ and $p$ is not a winner of the $\mathcal{E}$ election $(C', V)$. (As always, although we write "$V$" here, what we mean by $(C', V)$ is the election system is given each vote reduced down to just the candidates in $C'$.) Why? As to the "if" direction, if there exists such a subset $C'$, then the partition $(C', C - C')$ eliminates $p$ in a first-round election. As to the "only if" direction, suppose it is possible to ensure by some partition $(C_1, C_2)$ that $p$ is not an overall winner. If, under partition $(C_1, C_2)$, $p$ loses her first-round election, then whichever part of $(C_1, C_2)$ contained $p$ is a $C'$, $\{p\} \subseteq C' \subseteq C$, such that $p$ is not a winner of the $\mathcal{E}$ election $(C', V)$, which is what we were seeking. On the other hand if, under a partition $(C_1, C_2)$, $p$ survives the first round but loses in the second round, and the candidates in the second round are $D$, then $D$ is a $C'$ of the form we seek. This completes the proof of our "if and only if" characterization of on which instances destructive control by run-off partition of candidates, ties promote, can succeed. We note in passing that it follows that whenever success (eliminating $p$) is possible, success can be achieved even in the first round: in the "$D$" case above, we could use $(D, C - D)$ as our partition, rather than $(C_1, C_2)$, and $p$ would lose in the first round.

Now let us consider DC-PC-TP. Here, keeping $p$ from winning in a given instance can be accomplished if and only if there exists a set $C'$, $\{p\} \subseteq C' \subseteq C$, such that $p$ is not a winner of the $\mathcal{E}$ election $(C', V)$. This follows by precisely the same argument as above, except now the two first-round groups are not symmetric—one group has a first-round contest and one gets a first-round bye—and so in both directions of the proof we must use $p$-destroying collections as the side that has a first-round contest, e.g., in the $D$ part of the "only if" direction, we would use the partition that gives $C - D$ a first-round bye and has the candidates in $D$ participating in the first-round election (and eliminating $p$ there). But wait: The "if and only if" characterization we just got for which inputs to the DC-PC-TP problem lead to success is *exactly the same* as the characterization we got for the DC-RPC-TP problem. It follows, given that their input syntax is identical, that these two control types are not distinct but rather for all election systems are (and always have been) identical. This completes our proof of the first part of Theorem 3.8, but please keep not just that result in mind but also the shared characterization, as we centrally use it in showing our sample parts of Theorem 3.6 in Section 3.2.

We note in passing that if one were to—and we do *not* suggest doing so—define a new control type called "destructive control by unlimited deleting of candidates" (defined by asking if there exists a set $C'$, $\{p\} \subseteq C' \subseteq C$, for which under $\mathcal{E}$ $p$ is not a winner of $(C', V)$), that control type would also be identical to both DC-RPC-TP and DC-PC-TP: That is what the characterization proves. As a historical note, and to give credit where credit is due, the two characterizations we just gave, and the analogous two that we are about to give for the TE cases, exist in the literature, namely, they are used in passing inside a certain proof in Faliszewski et al. [FHHR09a], in particular, in the proof that the election systems "Copeland$^{\alpha}$" are "vulnerable" to DC-PC-TE/DC-PC-TP/DC-RPC-TE/DC-RPC-TP. So that paper already had the smoking gun in front of it—it simply didn't take the final step of realizing that identical characterizations mean the types collapse into identity, perhaps because the paper was looking at one election family, although its characterization certainly holds in general. But certainly, Theorem 3.8 does follow from those observations inside that proof of [FHHR09a].

As to the second part of Theorem 3.8, it is easy to see—and one can also find this within the above-cited "vulnerability" proof from [FHHR09a]—that for each of the DC-RPC-TE problem and the DC-PC-TE problem, $p$ can be prevented from winning the overall election if and only if there exists a $C'$, $\{p\} \subseteq C' \subseteq C$, such that in the $\mathcal{E}$ election $(C', V)$ either there are no winners, or there are two or more winners, or there is one winner and that winner is not $p$ (that is, $p$ is not a unique winner). The argument is analogous to the above one. And again, identical characterizations mean identical sets, and so the second part of Theorem 3.8 is proven. Again, the characterization is important too, as we use it in proving parts of Theorem 3.6.[11] ❑

[11] All the results of our paper are in the nonunique-winner model, which we feel is the more natural and attractive model. However, in light of our claim that four of the standard types pairwise collapse (in the nonunique-winner model), it seems important for the clarity of the literature to address the issue of whether the same claim holds in the unique-winner model. After all, all previous papers have blithely given separate proofs for cases that might be identical. What holds is the following: In the unique-winner model, one of the pairs still collapses and the other provably does not collapse.

# 5 Related Work

Since the seminal papers on the complexity of manipulation [BTT89a,BO91], bribery [FHH09], and control [BTT92], there has been a great deal of research trying to more broadly understand the algorithmic and complexity issues of those manipulative attacks—see the surveys [FHHR09b, FHH10] for an overview, and see also the outstanding paper [CSL07]. Although to the best of our knowledge search versus decision has not previously been a focus area in this line of work, the detailed analysis of the complexity of different attacks on particular systems has been a focus area. For example, detailed classifications of the complexities of the many types of constructive and destructive control actions on specific systems can be found in such work as [FHHR09a,ENR09, ER10,EPR11,BEH$^+$10,Men10]. These papers are about specific systems. In contrast, our "search reduces to decision" results hold for all systems. Our "if $\mathrm{P} \neq \mathrm{NP} \cap \mathrm{coNP}$" results on the other hand use that complexity-theoretic hypothesis to build specific systems that make decision easy while making search hard.

Existing papers that give polynomial-time attack algorithms against specific systems typically do so by (at least implicitly) finding a polynomial-time solution to the search problem. Probably the definition most related to the interests of this paper is the definition of "certifiably vulnerable" of Hemaspaandra, Hemaspaandra, and Rothe [HHR07], which captures the notion of demanding that an attack provide a successful action when one exists. That paper actually adds an "optimality" twist to that notion, but subsequent papers (e.g., [FHH11b,FHHR11]) when using the notion of certifiability take it to mean providing some successful action when one exists, rather than the

---

**Theorem 4.1**

1. DC-RPC-TP-UniqueWinner $\neq$ DC-PC-TP-UniqueWinner.
2. DC-RPC-TE-UniqueWinner $=$ DC-PC-TE-UniqueWinner.

The second part follows because those two types again clearly—and this too is in that same vulnerability proof of [FHHR09a]—have the same characterization as each other, namely, destructive control against $p$ is possible exactly if there is a set $C'$, $\{p\} \subseteq C' \subseteq C$, such that in the election $(C', V)$ either there are zero or more than one winners or there is one winner and that one is not $p$. Since this is the same characterization used in the second part of Theorem 3.8, we can even claim—although stating equality between nonunique-winner model items and unique-winner model items is unusual, they are all just sets and so this is mathematically meaningful—DC-RPC-TE $=$ DC-PC-TE $=$ DC-RPC-TE-UniqueWinner $=$ DC-PC-TE-UniqueWinner.

As to the first part, we can show that equality fails by a very simple example. Consider the following election system (which even has the property that its winner problem is in polynomial time, so the separation of these control types clearly does not inherently require election systems with hard winner problems): Whenever there is exactly one candidate in an election, that candidate is a winner. Whenever there are exactly two candidates, the candidate named Alice will be a winner if present, otherwise there are no winners. Whenever there are exactly three candidates and Alice is one of them, then Alice and the lexicographically smaller of the other two candidates are winners; in all other three-candidate elections no one wins. In all cases with four or more candidates, no one wins. Consider the input in which the candidates are Alice, Bob, and Carol (the voters are irrelevant to this example, so we don't bother to specify anything about them). For our stated election system, the input will be in the set DC-PC-TP-UniqueWinner, as we can make Alice not be a unique winner by giving all three candidates a first-round bye. This results in Alice and Bob being winners of the final election, and since Alice is thus not a unique winner the control action was successful. However, this instance is not in the set DC-RPC-TP-UniqueWinner, as no matter what partition is chosen it is not possible to prevent Alice from being a unique winner of the final election. We have thus given an election system that has an input that is a member of DC-PC-TP-UniqueWinner but not of DC-RPC-TP-UniqueWinner, and this suffices to show that the sets are not always equal.

The reader may wonder why in Theorems 3.8 and 4.1 we have three collapses and one separation. What is so special about the TP-UniqueWinner case that it evades the same type of shared characterization as the others? The answer is that that is the only one of the pairs in which the restrictiveness of the handling of the second round's output (only seeking a unique winner) exceeds the restrictiveness of the handling of the first round's output (all winners move forward), and this is precisely what our counterexample exploits.

"smallest in size/effort/cost" such action.

All our search reduces to decision results of course hold on all inputs. However, our "$\mathrm{P} \neq \mathrm{NP} \cap \mathrm{coNP}$"-induced results put decision versions in P while ensuring that their search versions are not polynomial-time computable. That latter part is a worst-case claim. It says that each polynomial-time algorithm for the search problem fails on some instance. It is easy to see from this that all polynomial-time algorithms must fail on infinitely many instances (if not, one would be able to patch with a finite table). But that leaves open the possibility that that infinite set of instances might be of very low density. That possibility is a real one, but we claim that it is intimately connected to the issue of the hardness-density of sets realizing (if $\mathrm{P} \neq \mathrm{NP} \cap \mathrm{coNP}$) the Borodin-Demers Condition. Since our constructions build on top of such sets, and our constructions turn each (well-formed) instance for a given, fixed set from the Borodin-Demers Condition into distinct and at most polynomially longer instances of the given manipulative action, it is easy to see that give or take the effect of the polynomial reduction's spreading effect, the density of hardness of our manipulative-action search problems can be made the same as the density of hardness of a set realizing the Borodin-Demers Condition.[12] In particular, if $f$ is any nondecreasing function, and for some Borodin-Demers Condition set $B$ it holds that every polynomial-time solution-finding (i.e., in this context, satisfying-assignment finding) algorithm errs, at infinitely many lengths $n$, on at least $f(n)$ of the strings in $B$ up to that length, then for each manipulative action (that appears in our $\mathrm{P} \neq \mathrm{NP} \cap \mathrm{coNP}$ theorems) there will exist an $\epsilon' > 0$ and an election system having a polynomial-time winner problem such that each search algorithm for that manipulative action with respect to that election system will err, at infinitely many lengths $n$, on at least $f(n^{\epsilon'})$ of the strings up to that length, but the decision problem will be in P. For example, if for some $\epsilon > 0$ and for some Borodin-Demers Condition set $B$ it holds that every polynomial-time solution-finding algorithm errs, at infinitely many lengths $n$, on at least $2^{n^{\epsilon}}$ of the strings in $B$ up to that length, then for each manipulative action (that appears in our $\mathrm{P} \neq \mathrm{NP} \cap \mathrm{coNP}$ theorems) there will exist an $\epsilon' > 0$ and an election system having a polynomial-time winner problem such that each search algorithm for that manipulative action with respect to that election system will err, at infinitely many lengths $n$, on at least $2^{n^{\epsilon'}}$ of the strings up to that length, but the decision problem will be in P. In short, although our paper speaks in terms of keeping search algorithms out of polynomial time, its proof infrastructure is enough to strongly address the issue of how often failure occurs—or at least to strongly link that to the open issue of how densely hard Borodin-Demers sets can be.

And we can extend the link from the previous paragraph back far further, all the way to achieving a connection to the hardness density of sets in $(\mathrm{NP} \cap \mathrm{coNP}) - \mathrm{P}$ (note: one might instead speak of $\mathrm{NP} \cap \mathrm{coNP}$, since sets in P are never hard). In particular, if one does a careful proof of the Borodin-Demers Theorem (see the proof in the course notes [Hem07]), one can ensure that the proof, from a fixed set $A$ in $(\mathrm{NP} \cap \mathrm{coNP}) - \mathrm{P}$, in effect gives a polynomial-time metric reduction to the satisfying-assignment question regarding a Borodin-Demers set (to be more specific, it creates a Borodin-Demers set $B$, and a polynomial-time function $g$, such that for all $x \in \Sigma^\star$ it holds that $g(x) \in B$, and for each $x$ it holds that $x \in A$ if and only if the first bit of some (equivalently, in the context of the properties of those $B$ created in the proof of the Borodin-Demers Theorem, every) satisfying assignment of satisfiable formula $g(x)$ has 1 as its first bit), and that, crucially, the function $g$ is injective (i.e., one-to-one). From that, and the obvious fact that the (polynomial-time) reduction function $g$ increases the length of its inputs at most polynomially, we can make the analogous conclusion to that made above: If $f$ is any nondecreasing function, and for some set $A \in \mathrm{NP} \cap \mathrm{coNP}$ it holds that every polynomial-time membership-in-$A$-testing algorithm errs, at infinitely many lengths $n$, on at least $f(n)$ of the strings up to that length, then there exists a Borodin-Demers set $B$ and an $\epsilon > 0$ such that each polynomial-time solution-finding (i.e., satisfying-assignment finding) algorithm errs, at infinitely many lengths $n$, on at least $f(n^{\epsilon})$ of the strings in $B$ up to that length. Combining the work of this paragraph and the previous one, we have the following general result, and specific corollary, showing that the density of hardness of the most densely hard sets in $\mathrm{NP} \cap \mathrm{coNP}$ (with some candidates for such often-hard sets being, for example, problems related to factoring) is inherited by our constructions, give or take an $\epsilon$ of flexibility. This provides a quantitative result regarding inheritance of frequency of hardness.

[12] By "each (well-formed) instance for a given, fixed set from the Borodin-Demers Condition" we mean strings $x \in B$, where $B$ is a fixed set $B$ in the sense of Definition 3.4, and the challenge is to find a satisfying assignment of $x$. We are not speaking of strings $x$ such that $x \notin B$. Since $B \in \mathrm{P}$, strings with $x \notin B$ are easy to detect, and so any polynomial-time solution-seeking algorithm for $B$ can easily be modified to never err on strings $x \notin B$. This is important to note, so as to see that the hardness inheritance claims later in the paragraph are valid (note in those the key related phrase "of the strings in $B$").

**Theorem 5.1** *If $f$ is any nondecreasing function, and for some set $A \in \mathrm{NP} \cap \mathrm{coNP}$ it holds that every polynomial-time membership-in-$A$-testing algorithm errs, at infinitely many lengths $n$, on at least $f(n)$ of the strings up to that length, then for each manipulative action (that appears in our $\mathrm{P} \neq \mathrm{NP} \cap \mathrm{coNP}$ theorems) there will exist an $\epsilon > 0$ and an election system having a polynomial-time winner problem such that each search algorithm for that manipulative action with respect to that election system will err, at infinitely many lengths $n$, on at least $f(n^{\epsilon})$ of the strings up to that length, but the decision problem will be in* P.

**Corollary 5.2** *If $f$ is any nondecreasing function, and for some set $A \in \mathrm{NP} \cap \mathrm{coNP}$ and some $\epsilon > 0$ it holds that every polynomial-time membership-in-$A$-testing algorithm errs, at infinitely many lengths $n$, on at least $2^{n^{\epsilon}}$ of the strings up to that length, then for each manipulative action (that appears in our $\mathrm{P} \neq \mathrm{NP} \cap \mathrm{coNP}$ theorems) there will exist an $\tilde{\epsilon} > 0$ and an election system having a polynomial-time winner problem such that each search algorithm for that manipulative action with respect to that election system will err, at infinitely many lengths $n$, on at least $2^{n^{\tilde{\epsilon}}}$ of the strings up to that length, but the decision problem will be in* P.

The analogous hardness inheritance argument works just as well for the case of frequent hardness being required to hold "almost everywhere," i.e., at all but a finite number of lengths. We thus also have the following claims.[13]

[13]Essentially the same argument shows that if there exists an NP set that is frequently hard then every set that is NP-hard with respect to polynomial-time, injective (i.e., one-to-one) reductions is roughly as hard, give or take the error frequency lower bound of $f(n)$ changing, as in the above four theorems and corollaries, to $f(n^{\epsilon})$. And since by the work of Berman and Hartmanis [BH77] essentially all familiar, natural NP-complete sets come so well-equipped with so-called padding functions that it holds that all NP sets reduce to them by polynomial-time injective reductions, this says that if any NP set is frequently hard, then essentially all familiar, natural NP-complete sets are almost as often hard, give or take the way the above $\epsilon$ transforms the argument (in effect shifting the argument to some root, possibly fractional, of itself). How can one harmonize this with the belief of many people that many or most natural NP-complete sets are not too frequently hard? One answer would be in that $\epsilon$ transformation. For example, suppose one suspects that there are NP sets relative to which all P heuristics are asymptotically wrong about half the time. (We find that a very natural conjecture. We mention in passing that the work of [HZ96] implies that there is an oracle relative to which that holds, and indeed that it holds with probability one relative to a random oracle; to see that from their "$\mathrm{NP}^A$ is $\mathrm{P}^A$-balanced-immune with probability one relative to a random oracle" result, simply note that their probability one P-balanced-immunity claim for NP when applied to each $\mathrm{P}^A$ set and its (also $\mathrm{P}^A$, of course) complement gives, combining the two "limit 1/2" statements those give from the definition of balanced immunity, that the limit of the density of the symmetric difference of each $\mathrm{P}^A$ set and the $\mathrm{NP}^A$ set is about 1/2.) Even that belief that NP totally confounds P heuristics will, under the "$\epsilon$," transform to an inherited error frequency lower bound, for familiar, natural NP-complete sets, of around $2^{n^{\epsilon}}$, with the $\epsilon > 0$ itself varying based on the particular familiar, natural NP-complete problem. That is what we suspect is the case. Others of course may resolve this issue by denying the hypothesis, i.e., by suspecting that no NP sets at all are too frequently hard.

**Theorem 5.3** *If $f$ is any nondecreasing function, and for some set $A \in \mathrm{NP} \cap \mathrm{coNP}$ it holds that every polynomial-time membership-in-A-testing algorithm errs, at almost every length $n$, on at least $f(n)$ of the strings up to that length, then for each manipulative action (that appears in our $\mathrm{P} \neq \mathrm{NP} \cap \mathrm{coNP}$ theorems) there will exist an $\epsilon > 0$ and an election system having a polynomial-time winner problem such that each search algorithm for that manipulative action with respect to that election system will err, at almost every length $n$, on at least $f(n^{\epsilon})$ of the strings up to that length, but the decision problem will be in* P.

**Corollary 5.4** *If $f$ is any nondecreasing function, and for some set $A \in \mathrm{NP} \cap \mathrm{coNP}$ and some $\epsilon > 0$ it holds that every polynomial-time membership-in-A-testing algorithm errs, at almost every length $n$, on at least $2^{n^{\epsilon}}$ of the strings up to that length, then for each manipulative action (that appears in our $\mathrm{P} \neq \mathrm{NP} \cap \mathrm{coNP}$ theorems) there will exist an $\tilde{\epsilon} > 0$ and an election system having a polynomial-time winner problem such that each search algorithm for that manipulative action with respect to that election system will err, at almost every length $n$, on at least $2^{n^{\tilde{\epsilon}}}$ of the strings up to that length, but the decision problem will be in* P.

On the complexity side of things, we use the Borodin-Demers Theorem, which gives a consequence of $\mathrm{P} \neq \mathrm{NP} \cap \mathrm{coNP}$ (building on Valiant [Val76] the work of Borodin and Demers appears as [BD76], see the discussion footnoting Theorem 3.3 for a more detailed history). Although we use the Borodin-Demers Theorem within the proofs that when $\mathrm{P} \neq \mathrm{NP} \cap \mathrm{coNP}$ search does not reduce to decision for various manipulative actions, our results showing that search does reduce to decision for many manipulative actions are related to a broad theme of complexity theory and of the Borodin-Demers paper [BD76], namely self-reducibility [MP79,Sch76,Val76,BD76]. The papers in that stream were very much interested—see especially [Val76,BD76]—in whether a function problem could be harder than its related decision problem. Cases when that cannot hold can often be shown by noting some self-reducibility structure in a problem. For example, for SAT, it has been known for decades that given a black box for the decision problem one can in polynomial time obtain a satisfying assignment when one exists, simply by walking appropriately down the tree created by the fact that SAT is so-called 2-disjunctively self-reducible. When showing that search reduces to decision we wish our results to hold for all election systems, not just those with polynomial-time winner problems, and this makes our algorithms slightly less transparent than the SAT example just given, as to how they are exploiting the structure of their problems, but they indeed are exploiting the structure very similarly. For example, in destructive control by run-off partition of candidates in the ties-promote model, in showing that search reduces to decision we (for reasons that come from the proofs regarding that) try to find a subcollection of candidates in which the hated candidate is not a winner, and we do so when such exists by finding a minimal such set, and we do that by from the input's starting point noting that if such a set exists either it is the entire input candidate set (which we cannot test directly as we cannot assume our winner problem is in P) or is a subset of some "after deleting one candidate other than the hated candidate"-adjustment to our current set; and so, given a decision-version black box, we argue that if a candidate set including the hated one gets the answer "Yes" and every way of deleting one of the nonhated candidates gets the answer "No," then the candidate set will solve the search problem, and further this gives us an iterative scheme for solving the search problem given access to the decision problem. (For the voter/candidate addition/deletion problems the lack of the assumption of a polynomial-time winner problem can be even more easily bypassed in the presence of a black box for the decision version, as for those problems the winner problem itself many-one polynomial-time reduces to the problem's decision version.) There is far too large a literature exploring the many aspects of search versus decision to cite it all here, but as an indication of how broad the literature is we mention a paper related to search versus decision for nondeterministic exponential-time computations [IT89] and a paper related to P-selectivity and self-reducibility [HNOS96]. Of course, the present paper is looking at concrete cases of search versus decision, in the context of manipulative actions on elections.

Does $\mathrm{P} \neq \mathrm{NP} \cap \mathrm{coNP}$ hold? Three closely related (to each other) old papers suggest that P versus $\mathrm{NP} \cap \mathrm{coNP}$ may be a hard issue to resolve [BI87,Tar89,HH90]. Nonetheless, there are a number of problems that are known to belong to $\mathrm{NP} \cap \mathrm{coNP}$ yet that despite intense effort have not been shown to belong to P. Such problems include important questions about lattice problems [GG00,AR05], stochastic games [Con92], parity games [Jur98], and factoring. Regarding the latter, it is well known that if $\mathrm{P} = \mathrm{NP} \cap \mathrm{coNP}$ then integer factoring is in polynomial time; this is to many people very strong evidence that $\mathrm{P} \neq \mathrm{NP} \cap \mathrm{coNP}$ (see [Kin10]). *Note that, thus, if one believes factoring is hard, then by our results one must also believe that search and decision differ in complexity for many types of manipulative attack.*

# 6 Conclusions

Papers on the complexity of manipulative attacks on elections have typically defined their concepts in terms of decision problems while carrying out actual polynomial-time attacks, as a practical matter, through algorithms targeting the search versions. This naturally raises the question of whether simple search versions and simple decision problems inherently stand or fall together. This paper addresses that question for all the standard types of manipulative attacks. Interestingly, there is not a single answer. Rather, for many types of attacks, we proved that for all election systems search reduces to decision, and so their complexity stands or falls together. But for many other types of manipulative attacks, we showed that if $\mathrm{P} \neq \mathrm{NP} \cap \mathrm{coNP}$, then there are election systems, having polynomial-time winner problems, for which decision is easy but search is hard. Table 1 summarizes our results. For the four types of partition related to candidates, we saw an interesting behavior, namely, that the constructive and destructive cases differed in their relationship between search and decision.

Since our paper shows that one cannot simply assume that search and decision problems stand together, the natural lesson to draw is that in framing definitions and questions, heightened attention should in the future be given to search versions.